\documentclass[onecolumn,10pt]{revtex4-1}

\usepackage{graphicx}
\usepackage{amsmath}
\DeclareMathOperator\erf{erf}
\renewcommand{\arraystretch}{1.8}

\begin{document}

\title{Transforming Gaussian correlations. Applications to generating long-range power-law correlated time series with arbitrary distribution}
\author{Pedro Carpena}
\email{pcarpena@ctima.uma.es}
\author{Pedro A. Bernaola-Galv\'an}
\author{Manuel G\'omez-Extremera}
\author{Ana V. Coronado}
\affiliation{Departamento de F\'\i sica Aplicada II, E.T.S.I. de Telecomunicaci\'on, Universidad de M\'alaga. 29071, M\'alaga, Spain.}

\begin{abstract}
The observable outputs of many complex dynamical systems consist in time series exhibiting autocorrelation functions of great diversity of behaviors, including
long-range power-law autocorrelation functions, as a signature of interactions operating at many temporal or spatial scales.
Often, numerical algorithms able to generate correlated noises reproducing the properties of real time series
are used to study and characterize such systems. Typically, those algorithms produce \textsl{Gaussian} time series. However, real,
experimentally observed time series are often non-Gaussian, and may follow distributions with a diversity of 
behaviors concerning the support, the symmetry or the tail properties. Given a correlated Gaussian time series, it is always possible to transform it into 
a time series with a different distribution, but the question is how this transformation affects the behavior of the autocorrelation function. Here, we study analytically and numerically
how the Pearson's correlation of two Gaussian variables changes when the variables are transformed to follow a different destination distribution. Specifically, we consider
bounded and unbounded distributions, symmetric and non-symmetric distributions, and distributions with different tail properties, from decays faster than exponential
to heavy tail cases including power-laws, and we find how these properties affect the correlation of the final variables. We extend these results to Gaussian time series which are transformed
to have a different marginal distribution, and show how the autocorrelation function of the final non-Gaussian time series depends on the Gaussian correlations and on the final marginal distribution.
As an application of our results, we propose how to generalize standard algorithms producing Gaussian power-law correlated time series in order to create synthetic time series with arbitrary distribution and controlled power-law correlations. Finally, we show a practical example of this algorithm by generating time series mimicking the marginal distribution and the power-law tail of the autocorrelation function of a real time series: the absolute returns of stock prices.

\end{abstract}

\maketitle
\section{Introduction}

The observable outputs of a great diversity of dynamical systems consist in correlated time series, and the corresponding autocorrelation functions may have many different functional forms, characterizing
the underlying dynamics which is tipically not explicitly known. For example, when the dynamics presents a characteristic time (or scale), exponentially decreasing autocorrelations functions are found. Also, when dealing with complex systems with interactions working at many time or spacial scales, then very often one finds time series with long-range, power-law decaying autocorrelation functions. 
Such time series can be found, for example, in Physiology (heartbeat dynamics \cite{Peng93}, brain activity \cite{correlations_brain}, respiration \cite{respiration}, postural control \cite{postural_control,maite}), Biology (DNA and protein sequences) \cite{Peng_1992,voss_dna_1992}, Economics (stock market activity) \cite{economy}, Music \cite{voss_music}, Meteorology (temperature or rain precipitation) \cite{meteo,geo01}, Geophysics (seismic signals \cite{seismic}), and many other fields. Although there exist other measures of dependence between variables, the analysis of the linear correlations as measured by the autocorrelation function is important since, in many cases only the linear correlations are considered to quantify the complexity and the scaling properties of natural time series, e.g. heart-rate variability \cite{heart-rate-review2006} or financial time series \cite{financial-wavelets2001}.

Indeed, no matter the functional form of the autocorrelation function of the observable time series, a common approach to study and characterize the underlying dynamical system is often based on numerical algorithms able to generate surrogate signals replicating the correlation properties of the real time series. For example, autoregressive processes of order 1 (AR(1)) are able to produce
time series with exponentially decreasing correlations. As another example, the Fourier Filtering method (FFM) \cite{FFM1,FFM2} is probably the most successful algorithm able to produce power-law correlated time series. Other techniques are able to generate surrogate signals with the same autocorrelation function as the experimental time series: they obtain the power spectrum 
of the real signal, and modify the Fourier phases without altering the power spectrum values (see \cite{theiler} and later generalizations \cite{schreiber,kugiumtzis,keylock}, and applications, for
instance, to climatic records \cite{halley}). When the signal is Fourier-transformed back into time domain, the autocorrelation function is identical to that of the real signal due to the Wiener-Khinchin theorem. 

However, most of such algorithms produce \textsl{Gaussian} time series. In contrast, the marginal probability distribution of real-world correlated time series are often non-Gaussian \cite{Peng93}: Indeed, the distributions can exhibit very different behaviors, concerning the support (bounded or unbounded), the symmetry (symmetric or not) or the tail behavior (exponential decay, faster than exponential, heavy tail, etc). As a consequence, once the Gaussian output of any given algorithm is available, a final transformation is required to change the Gaussian distribution to the desired final marginal distribution of the experimental time series. Nevertheless, this last transformation \textsl{always} modifies the autocorrelation function of the Gaussian time series (identical to that of the real time series), so that the final time series follows the same marginal distribution as the real one but with a different autocorrelation function. Although this drawback may be reduced
by iterative procedures (as the one in \cite{schreiber}), the difference between both the Gaussian and the final autocorrelation functions can be specially dramatic when the final marginal distribution is far from Gaussian.

Formally speaking, let us consider a Gaussian time series $\{z_{G,i}\}$, $i=1,2,\ldots,N$, with an autocorrelation function $C_G(\ell)$ given by:
\begin{equation}
C_G(\ell)\equiv \frac{\langle z_{G,i} z_{G,i+\ell}\rangle -\mu_G^2}{\sigma_G^2} \label{def-cor}
\end{equation}
where $\mu_G$ and $\sigma_G^2$ are the mean and the variance of the Gaussian distribution respectively, and the subscript $G$ refers to Gaussian distribution from now on. 
Without loss of generality, the time series $\{z_{G,i}\}$ can be normalized to have 0 mean and unit standard deviation, so that $z_{G,i} \sim {\cal N}(0,1)$  $\forall i$, and then
$C_G(\ell)\equiv \langle z_{G,i} z_{G,i+\ell}\rangle$. In this case, the corresponding probability
density $\varphi(z_G)$ and cumulative distribution $\Phi(z_G)$ are given by: 
\begin{eqnarray}
\varphi(z_G)&=&\frac{1}{\sqrt{2\pi}} e^{-\frac{z_G^2}{2}} \nonumber\\
\Phi(z_G)&=&\int_{-\infty}^{z_G} \frac{1}{\sqrt{2\pi}} e^{-\frac{\xi^2}{2}}\, d\xi =
\frac{1}{2}\left[\erf\left(\frac{z_G}{\sqrt{2}}\right) +1 \right] \label{fi}
\end{eqnarray}
with $\erf(x)$ the standard error function.

We note that $\{z_{G,i}\}$ can be transformed into a time series $\{z_i\}$ following any arbitrary \textsl{destination} marginal distribution characterized by a probability density $f(x)$ and a cumulative distribution $F(x)$ by using the standard technique of the inverse cumulative distribution \cite{numrec}:
\begin{equation}
z_i=F^{-1}[\Phi(z_{G,i})], \quad i=1,2,\ldots,N \label{transforma1}
\end{equation} 
with $F^{-1}(x)$ the inverse cumulative distribution of the destination distribution. It is well known that strictly increasing transformations like (\ref{transforma1}), when applied to a pair of random variables such as $z_{G,i}$ and $z_{G,i+\ell}$, do not modify the dependence between them \cite{copulas99} and therefore the corresponding values of the final time series, $z_i$ and $z_{i+\ell}$, preserve their statistical dependence as measured by rank statistics like Kendall's $\tau$ and Spearman's $\rho$ \cite{copulas99}, or by Information Theory functionals like mutual information \cite{mi90}. Nevertheless, the transformation (\ref{transforma1}) does \textsl{not} preserve the linear correlations, as measured by the autocorrelation function. The autocorrelation of the original Gaussian series
$\{z_{G,i}\}$ is $C_G(\ell)$ (\ref{def-cor}), and after the transformation, the autocorrelation function $C(\ell)$ of the final time series $\{z_i\}$ is defined as  
\begin{equation}
C(\ell)\equiv \frac{\langle z_i z_{i+\ell}\rangle -\mu^2}{\sigma^2} \label{cor-nongau}
\end{equation}
with $\mu$ and $\sigma^2$ the mean and variance of the arbitrary destination probability density $f(x)$.  The transformation changes the correlations, so that $C(\ell) \neq C_G(\ell)$, and in general, the behavior of $C(\ell)$ depends on the final marginal distribution.

In this work, we study how the autocorrelation function of a given Gaussian time series changes when the marginal distribution is modified from Gaussian
to the desired final distribution. This problem has been intensively investigated in several contexts \cite{theiler,schreiber,kugiumtzis,keylock,halley,li75,chen01,kugi10,cario}, but we focus here on how $C(\ell)$ depends on $C_G(\ell)$ for a diversity of destination probability distributions with fundamentally different statistical properties (support, symmetry and tail behavior), in order to find out which properties of the final distribution control the behavior of $C(\ell)$, and the differences between $C(\ell)$ and $C_G(\ell)$.

In addition, we also study under which conditions the asymptotic properties of $C_G(\ell)$ (for large $\ell$) are preserved after the transformation (\ref{transforma1}), so that $C(\ell)$ exhibits the same
asymptotic behavior as $C_G(\ell)$. This is of special relevance when studying time series with fractal, long-range power-law decaying autocorrelation functions, that appear ubiquitously in many complex dynamical systems \cite{Peng93,correlations_brain,respiration,postural_control,maite,Peng_1992,voss_dna_1992,economy,voss_music,meteo,geo01,seismic} as we mentioned above, and that are typically non-Gaussian. Note that the FFM algorithm \cite{FFM1,FFM2} produces Gaussian time series with controlled, power-law behaved $C_G(\ell)$. When the transformation (\ref{transforma1}) preserves the asymptotic behavior of $C_G(\ell)$, we can generalize the FFM algorithm to produce time series with arbitrary distribution and with autocorrelation function $C(\ell)$ with the same controlled power-law behavior. 

The paper is organized as follows: In Sec. II we consider two Gaussian variables with a linear correlation value given by $C_G$, and we transform them using (\ref{transforma1}) into two non-Gaussian variables with arbitrary marginal distribution and with linear correlation given by $C$, and obtain some general properties of the $C(C_G)$ function. The specific results of the function $C(C_G)$ for several destination distributions with different statistical properties are shown in Sec. III. The extension of these results to time series is addressed in Sec. IV, where we also 
include two applications: 1) A generalization of the FFM algorithm able to synthesize generic power-law correlated time series with arbitrary marginal distribution; and 2) a practical example where
we generate a time series mimicking the distribution and the power-law tail of the autocorrelation function of a real-world power-law correlated time series: the absolute returns of the stock price
of a technological company. Finally, in Sec. V we present our conclusions.

\section{Transforming Gaussian correlations}

Let us consider a generic Gaussian time series $\{z_{G,i}\}$, and without loss of generality, let us also assume that $z_{G,i} \sim {\cal N}(0,1)$ $\forall i$, so that the corresponding probability density $\varphi(z_G)$ and cumulative distribution $\Phi(z_G)$ are given in Eq. (\ref{fi}).

For the sake of simplicity, in this section we work with the pair of variables $x_G$ and $y_G$ defined respectively as $x_G\equiv z_{G,i}$ and $y_G\equiv z_{G,i+\ell}$ (i.e. we omit the $\ell$ dependence), and therefore the linear correlation between $x_G$ and $y_G$ is given by $C_G$, i.e. $C_G=\langle x_G y_G \rangle$.  Similarly, given a destination marginal distribution characterized by a probability
density $f(x)$ and cumulative distribution $F(x)$, we use the transformation (\ref{transforma1}) to obtain the final variables $x$ and $y$ given by  
\begin{equation}
x=F^{-1}(\Phi(x_G)), \quad y=F^{-1}(\Phi(y_G)) \label{tr}
\end{equation}
or, in other words, we also define $x\equiv z_i$ and $y\equiv z_{i+\ell}$. The linear correlation $C$ between these two variables is then
\begin{equation}
C=\frac{\langle x y \rangle -\mu^2}{\sigma^2} \label{cxy}
\end{equation}  
with $\mu$ and $\sigma^2$ the mean and variance of destination marginal distribution.

The natural question is how $C$ depends on the Gaussian correlation $C_G$, $C(C_G)$, or, in other words, how  the correlation changes when the distributions change from Gaussian to an arbitrary destination probability density $f(x)$. Obviously, $C(C_G)$ will depend on the specific properties of $f(x)$. In this work, we have investigated several destination probability distributions which have been selected to reflect different fundamental statistical properties: (i) We have considered distributions with bounded and unbounded support; (ii) For the unbounded support case, we study examples of distributions with different tail behavior ranging from faster than exponential to heavy tail cases including power-law tail behavior; and (iii), we have also considered symmetric and non-symmetric probability distributions. In this context, symmetric means that there exists a 'central point' $x_0$ such that
$f(x_0+x)=f(x_0-x)$, from where it is easy to obtain that $x_0$ corresponds to the median $m$ and the mean $\mu$ of the distribution. For convenience, in Table I we show separately symmetric
and non-symmetric distributions because, as we will see later, both groups exhibit different behavior.

\begin{table}
\caption{Symmetric and  non-symmetric destination distributions considered in this work. We include the common name, the support, the probability density, the cumulative distribution and the
inverse cumulative distribution.}
\begin{tabular}{|c|c|c|c|c|}
\hline
\hline
\multicolumn{5}{|c|}{Symmetric distributions}\\
\hline \hline
name & support & $f(x)$ & $F(x)$ & $F^{-1}(y)$, $y\in \lbrack 0,1]$ \\ \hline\hline
uniform & $[a,b]$ & $\frac{1}{b-a}$ & $\frac{x-a}{b-a}$ & $a+(b-a)y$ \\ \hline
arcsine & $(a,b)$ & $\frac{1}{\pi \sqrt{(x-a)(b-x)}}$ & $\frac{2}{\pi }\arcsin (%
\sqrt{\frac{x-a}{b-a}})$ & $a+(b-a)\sin ^{2}(\frac{\pi }{2}y)$ \\ \hline
logistic & $(-\infty ,\infty )$ & $\frac{\exp (-(x-\mu )/s)}{s(1+\exp
(-(x-\mu )/s))^{2}},$ $s>0$ & $\frac{1}{1+\exp (-(x-\mu )/s)}$ & $\mu -s\ln
\left( \frac{1-y}{y}\right) $ \\ \hline
Laplace & $(-\infty ,\infty )$ & $\frac{1}{2\lambda }\exp (-\frac{|x-\mu|}{%
\lambda }),$ $\lambda >0$ & $\left\{  
\renewcommand{\arraystretch}{1.5}
\begin{tabular}{l}
$\frac{1}{2}e^{(x-\mu)/\lambda }$ if $x<\mu$ \\ 
$1-\frac{1}{2}e^{(-x+\mu)/\lambda }$ if $x\geq \mu$%
\end{tabular}%
\right.  $ & $\left\{ 
\renewcommand{\arraystretch}{1.5}
\begin{tabular}{l}
$\mu+\lambda\ln (2y)$ if $y<0.5$ \\ 
$\mu-\lambda\ln (2(1-y))$ if $y\geq 0.5$%
\end{tabular}%
\right. $ \\ \hline
\begin{tabular}{c}
Pareto \\ 
symmetric%
\end{tabular}
& $(-\infty ,\infty )$ & $\frac{\varepsilon}{2a}(1+\frac{\left\vert x-\mu\right\vert}{a} )^{-(\varepsilon+1)},$
$\varepsilon>2, a>0$ & $\left\{ 
\renewcommand{\arraystretch}{1.5}
\begin{tabular}{l}
$\frac{1}{2}(1+\frac{\mu-x}{a})^{-\varepsilon}$ if $x<\mu$ \\ 
$1-\frac{1}{2}(1+\frac{x-\mu}{a})^{-\varepsilon}$ if $x\geq \mu$%
\end{tabular}%
\right. $ & $\left\{ 
\renewcommand{\arraystretch}{1.5}
\begin{tabular}{l}
$\mu+a\left(1-(2y)^{-1/\varepsilon}\right)$ if $y<0.5$ \\ 
$\mu+a\left((2(1-y))^{-1/\varepsilon}-1\right)$ if $y\geq 0.5$%
\end{tabular}%
\right. $ \\ \hline \hline
\multicolumn{5}{|c|}{Non-symmetric distributions} \\ \hline \hline
name & support & $f(x)$ & $F(x)$ & $F^{-1}(y)$, $y\in \lbrack 0,1]$ \\ \hline\hline
exponential & $[0,\infty )$ & $\frac{1}{\lambda }e^{-x/\lambda }$, $\lambda
>0$ & $1-e^{-x/\lambda }$ & $-\lambda \ln (1-y)$ \\ \hline
Weibull & $(0,\infty )$ & $\frac{\delta }{\lambda }\left( \frac{x}{\lambda }%
\right) ^{\delta -1}\exp \left( -\left( \frac{x}{\lambda }\right) ^{\delta
}\right) ,$ \ $\lambda ,\delta >0$ & $1-\exp \left( -\left( \frac{x}{\lambda 
}\right) ^{\delta }\right) $ & $\lambda (-\ln (1-y))^{1/\delta }$ \\ \hline
lognormal & $(0,\infty )$ & $\frac{1}{\sqrt{2\pi }s x}\exp \left( -%
\frac{(\ln x-m )^{2}}{2s^{2}}\right) ,$ $ s >0$ & $\frac{1}{2}+%
\frac{1}{2}\erf\left( \frac{\ln x-m }{\sqrt{2}s }\right) $ & $%
\exp \left( m +\sqrt{2}s \erf^{-1}(2y-1)\right) $ \\ \hline
Pareto & $[0,\infty )$ & $\frac{\varepsilon}{a}(1+\frac{x}{a})^{-(\varepsilon+1)}$, $a>0, \varepsilon>2$ & $1-\left(\frac{a}{a+x}\right)^{\varepsilon}$ & $%
a\left((1-y)^{-1/\varepsilon}-1\right)$ \\ \hline
\end{tabular}
\end{table}

The problem of determining how $C$ depends on $C_G$ can be tackled as follows: First, we recall that the correlation $C_G$ between the Gaussian variables $x_G$ and $y_G$ is purely linear, and this is
equivalent to affirm that the joint probability density of the pair $(x_G,y_G)$ is the bivariate Gaussian distribution, $\varphi_2(x_G,y_G,C_G)$, given by:
\begin{equation}
\varphi_2(x_G,y_G,C_G)=\frac{1}{2\pi\sqrt{1-C_G^2}}\exp\left( -\frac{ x_G^2+y_G^2-2C_G x_G y_G }{2(1-C_G^2)}
 \right) \label{bivar}
\end{equation}
Second, given the destination probability density $f(x)$, the corresponding $\mu$ and $\sigma$ in (\ref{cxy}) are known. Then, the problem of calculating $C$ is reduced to obtaining $\langle x y \rangle$. As $x$ and $y$ depend functionally on $x_G$ and
$y_G$ respectively (Eq. \ref{tr}), and as the joint probability density of $x_G$ and $y_G$ is $\varphi_2(x_G,y_G,C_G)$ (Eq. (\ref{bivar})) then
\begin{equation}
\langle x y \rangle(C_G)=\int_{-\infty}^{\infty}\int_{-\infty}^{\infty} F^{-1}(\Phi(x_G))\, F^{-1}(\Phi(y_G))\, \varphi_2(x_G,y_G,C_G)\, dy_G \, dx_G \label{xy}
\end{equation}
where we have written explicitly the dependence of $\langle x y \rangle$ on $C_G$. The integral in (\ref{xy}) does not admit an analytical solution in general (except in two cases, discussed below). Therefore, it has to be solved numerically, and  after introducing $\langle x y \rangle (C_G)$ in  Eq. (\ref{cxy}), the final result $C(C_G)$ is obtained. A similar approach to the to the one we have presented above and the numerical solution of
the integral in Eq. (\ref{xy}) was firstly considered in \cite{li75} when addressing the problem of generating random numbers with specific marginal distribution and prescribed correlations, later extended
to $n$-dimensional vectors \cite{chen01}. Also, parametric solutions of the integral in Eq. (\ref{xy}) has been proposed \cite{kugi10}.

However, even without solving the integral in Eq. (\ref{xy}), some important properties of $C(C_G)$ can be inferred: 
\begin{enumerate}
\item[(i)] \textsl{The transformation (\ref{tr}) maps uncorrelated Gaussian variables into uncorrelated variables.}  Note that when the Gaussian variables are uncorrelated, $C_G=0$, then their joint probability distribution factorizes, $\varphi_2(x_G,y_G,0)=\varphi(x_G)\, \varphi(y_G)$. As a consequence, the double integral factorizes in the product of two identical integrals each one giving $\mu$, the mean of the destination distribution, and then $\langle x y \rangle(0)=\mu^2$. By inserting $\langle x y \rangle(0)$ in Eq. (\ref{cxy}) we get $C(0)=0$. 

\item[(ii)] \textsl{If $C_G>0$, then $C>0$. Also, if $C_G<0$ then $C<0$}. This fact is inherited from the properties of $\varphi_2(x_G,y_G,C_G)$: let $f(x_G,y_G)$ and $g(x_G,y_G)$ be two non-decreasing
functions of the Gaussian variables $x_G$ and $y_G$ with joint distribution  $\varphi_2(x_G,y_G,C_G)$. Then, the covariance ${\rm{Cov}}[fg]$ is positive for $C_G>0$ and negative for $C_G <0$ \cite{joint}. Considering $f(x_G,y_G)\equiv F^{-1}(\Phi(x_G))$ and $g(x_G,y_G)\equiv F^{-1}(\Phi(y_G))$, since by construction $F^{-1}(\Phi(x))$ is a non-decreasing function,  the property holds.

\item[(iii)] \textsl{The function $C(C_G)$ is non-decreasing for $C_G\in(-1,1)$}. This property is easy to prove for $C_G>0$, and we do it in Sec. \ref{series}. For $C_G<0$, a prove can be found in
\cite{cario}. 
 
\item[(iv)] \textsl{For symmetric destination marginal distributions $C(C_G)$ is and odd function.} In general, the transformation $x=F^{-1}(\Phi(x_G))$ (and similarly for $y$) converts trivially the median $m_G$ of the Gaussian density $\varphi(x)$ into the median $m$ of $f(x)$ since by definition $\Phi(m_G)=1/2$ and $F^{-1}(1/2)=m$. When $f(x)$ is symmetric, then $m=\mu$ and without loss of generality we can fix them to 0, i.e. $m=\mu=0$ as it happens with $\varphi(x)$ for which $m_G=\mu_G=0$. Let us consider a given $x_G$ value such that $x_G>m_G=0$. Then, $\Phi(x_G)=1/2+a> 1/2$ for some $a \in(0,1/2)$,  and then $x=F^{-1}(a+1/2)>F^{-1}(1/2)=0$. Therefore, a positive $x_G$ is transformed into a positive $x$. Similarly, if we consider $x'_G<m_G=0$, then $\Phi(x'_G)=1/2-b< 1/2$ for some $b \in(0,1/2)$, and therefore $x'=F^{-1}(\Phi(x'_G))=F^{-1}(1/2 -b)< F^{-1}(1/2)=0$, thus implying that a negative $x'_G$ is transformed into a negative $x'$. Finally, if we take $x_G$ and $-x_G$ and transform both, since $\varphi(x)$ is symmetric then $\Phi(x_G)=1/2 +a$ and $\Phi(-x_G)=1/2-a$ both with the same $a\in(0,1/2)$. But since $f(x)$ is also symmetric, if $F^{-1}(1/2 + a)=x$ then $F^{-1}(1/2 -a )=-x$. Altogether, we get that if $x_G$ is transformed into $x$ then $-x_G$ is transformed into $-x$, or formally speaking  $F^{-1}(\Phi(x_G))$ is an \textsl{odd} function. Now, consider the integral in Eq. (\ref{xy}) giving $\langle x y \rangle(C_G)$, and let us try to calculate  $\langle x y \rangle(-C_G)$, i.e. we invert the sign of the Gaussian correlation. But due to the form of $\varphi_2(x_G,y_G,C_G)$, we note that $\varphi_2(x_G,y_G,-C_G)=\varphi_2(-x_G,y_G,C_G)$, and as we have just shown that $F^{-1}(\Phi(-x_G))=-F^{-1}(\Phi(x_G))$, from Eq. (\ref{xy}) we get that $\langle x y \rangle(-C_G)=-\langle x y \rangle(C_G)$. Since $f(x)$ is symmetric with fixed $\mu=0$, we finally obtain that 
\begin{equation}
C(-C_G)=-C(C_G). \label{odd}
\end{equation}

\item[(v)] \textsl{The function $C(C_G)$ maps the interval $(-1,1)$ into $(C_{\min},1)$ with $-1 \leq C_{\min}<0$}. Note that $C_G=1$ implies that the Gaussian variables $x_G$ and $y_G$ are identical,
$x_G=y_G$. Then, they are transformed to identical non-Gaussian variables, $x=y$ so that $C=1$. The result for $C_G=-1$, i.e., $C(-1)=C_{\min}$ with $-1 \leq C_{\min}<0$ is proved in Sec. \ref{series}. The equality $C_{\min}=-1$ holds for symmetric marginal distributions since $C(-1)=-C(1)=-1$, while in general for non-symmetric marginal distributions $-1 < C_{\min}<0$.

\end{enumerate}

\subsection{Series expansion and approximate solution} \label{series}

As we have stated above, the 2D integral in Eq. (\ref{xy}) does not admit an analytical solution in general. Then, given a destination density $f(x)$, one has to solve it numerically for any value of $C_G$.
However, the 2D integral can be approximated by using a Taylor expansion in terms of $C_G$, which will be of interest in Sec. IV. Then, we can write
\begin{equation}
\langle x y \rangle(C_G)=\mu^2+\sum_{n=1}^{\infty}a_n C_G^n \label{serie1}
\end{equation}
where we have used that $\langle x y \rangle(0)=\mu^2$, and 
\begin{equation}
a_n=\frac{1}{n!}\left.\frac{d^n \langle x y \rangle}{d C_G^n}\right|_{C_G=0}=\frac{1}{n!}\int_{-\infty}^{\infty}\int_{-\infty}^{\infty} F^{-1}(\Phi(x_G))\, F^{-1}(\Phi(y_G))\, 
\left.\frac{\partial^n \varphi_2(x_G,y_G,C_G)}{\partial C_G^n}\right|_{C_G=0} \, dy_G \, dx_G \label{an}
\end{equation}
It is not difficult to check that when evaluating the partial derivatives of $\varphi_2(x_G,y_G,C_G)$ at $C_G=0$ then the result factorizes in terms of the one-variable Gaussian densities
$\varphi(x_G)$ and $\varphi(y_G)$:
\begin{equation}
\left.\frac{\partial^n \varphi_2(x_G,y_G,C_G)}{\partial C_G^n}\right|_{C_G=0} = \left[(-1)^n\frac{d^n \varphi(x_G)}{dx_G^n} \right] \left[ (-1)^n\frac{d^n \varphi(y_G)}{dy_G^n}\right]
\end{equation}
Therefore, the 2D integral in Eq. (\ref{an}) is given by the product of two 1D identical integrals on $x_G$ and $y_G$ or
\begin{equation}
a_n=\frac{1}{n!}\left(  \int_{-\infty}^{\infty} F^{-1}(\Phi(x_G))\, \left[(-1)^n\frac{d^n \varphi(x_G)}{dx_G^n} \right]\, dx_G \right)^2
\end{equation}
and, using the properties of the derivatives of the Gaussian density $\varphi(x_G)$,
\begin{equation}
(-1)^n\frac{d^n \varphi(x_G)}{dx^n} = H_n(x_G) \varphi(x_G)
\end{equation}
with $H_n(x)$ the $n$-th order Hermite's polynomial. Then, we finally get
\begin{equation}
a_n=\frac{1}{n!}\left(  \int_{-\infty}^{\infty} F^{-1}(\Phi(x_G))\, H_n(x_G)\, \varphi(x_G)\, dx_G \right)^2  \label{an1}
\end{equation}
After inserting the final expression for $a_n$ (\ref{an}) into Eq. (\ref{serie1}), then by introducing $\langle x y \rangle(C_G)$ into Eq. (\ref{cxy}) we finally obtain
\begin{equation}
C(C_G)=\sum_{n=1}^{\infty} b_n C_G^n \label{serie2}
\end{equation}
with 
\begin{equation}
b_n=\frac{a_n}{\sigma^2}=\frac{1}{n! \,\sigma^2} \left(  \int_{-\infty}^{\infty} F^{-1}(\Phi(x_G))\, H_n(x_G)\, \varphi(x_G)\, dx_G \right)^2 \label{bn}
\end{equation} 
Then, any coefficient $b_n$ is obtained in general by solving numerically a 1D integral. From the computational point
of view, it may be interesting  to consider $n$ terms in the expansion (\ref{serie2}) as an approximation to $C(C_G)$ thus
solving $n$ 1D integrals instead of the 2D integral in Eq. (\ref{xy}). In addition, the expansion coefficients $b_n$ present
some convenient properties:
\begin{enumerate}
\item[(i)] By definition in Eq. (\ref{bn}), all the coefficients $b_n$ are positive. Then, in most cases the convergence of the expansion (\ref{serie2}) 
is typically fast. Using that $b_n$ are positive, we can also prove that $C(C_G)$ is an increasing function for positive $C_G$: note that
\begin{equation}
\frac{dC}{dC_G}=\sum_{n=1}^{\infty} n b_n C_G^{n-1}
\end{equation}  
which is trivially positive for $C_G>0$.

\item[(ii)] The extreme value $C_G=1$ corresponds to the case $x_G=y_G$ since both stochastic variables are of ${\cal N}(0,1)$ type. Therefore, when we transform
$x_G$ and $y_G$ we obtain $x=y$ thus implying that $C(C_G=1)=1$ as we have shown above, and from Eq. (\ref{serie2}) we get  
\begin{equation}
\sum_{n=1}^{\infty} b_n =1 \label{serie3}
\end{equation}
Then, each coefficient $b_n$ corresponds to a normalized weight characterizing the contribution of the $n$-th term in the expansion (\ref{serie2}).
In this sense, the first coefficient $b_1$ is a measure of the linearity of the function $C(C_G)$ and will be important in Sec. \ref{gen-series}. 

\item[(iii)] As we have shown above, when the probability density $f(x)$ of the stochastic variables $x$ and $y$ is symmetric, then $C(C_G)$ is an odd function (Eq. (\ref{odd})). In such case, $b_n=0$ for $n$ even and only the odd terms are present in the expansion of Eq. (\ref{serie2}):
\begin{equation}
C(C_G)=\sum_{n=1}^{\infty} b_{2n-1} C_G^{2n-1} \label{serie4}
\end{equation}
Since for symmetric distributions $b_2=0$, for sufficiently small $C_G$ values $C(C_G)$ is essentially linear, with profound implications in the generation of power-law correlated time series as we see
in Sec. \ref{gen-series}.

\item[(v)] We can use the previous properties to prove that $C(-1)=C_{\min}$ with $-1\leq C_{\min} <0$.  We already know that $C(-1)<0$ since negative Gaussian correlations are mapped into negative correlations. In addition, from (\ref{serie2}) we get
\begin{eqnarray}
C(-1)&=&\sum_{n=1}^{\infty} b_n (-1)^n = \sum_{n=1}^{\infty} b_{2n} - \sum_{n=1}^{\infty} b_{2n-1} \nonumber\\
&\geq& -\sum_{n=1}^{\infty} b_{2n} - \sum_{n=1}^{\infty} b_{2n-1} = -\sum_{n=1}^{\infty} b_n =-1
\end{eqnarray}
thus completing the prove. Also, the equality $C_{\min}=-1$ is valid only if $b_{2n}=0$ $\forall n$, so that the expansion (\ref{serie2}) only contains odd terms, as in (\ref{serie4}). This implies
that $C(C_G)$ is an odd function, and therefore the final marginal distribution must be symmetric. Altogether, $C_{\min}=-1$ for symmetric distributions, and $-1<C_{\min}<0$ for non-symmetric ones.
\end{enumerate}

In addition to these general properties, the results of the specific behavior of the $C(C_G)$ function for the distributions shown in Table I are discussed in the next section. 

\section{Results for several distributions} \label{distribuciones}

After analyzing the general properties of $C(C_G)$, in this section we present the specific results of $C(C_G)$ for the symmetric and non-symmetric distributions in Table I. 
Apart from the symmetry, the criteria we have followed for selecting these examples are varied: First, we have tried to consider distributions found in real data with different fundamental
properties such as the support (bounded or unbounded) and, for the unbounded cases, the behavior of the tail of the distribution (exponential, faster decay than exponential and heavy-tail cases).
In addition, all the selected examples present an inverse cumulative distribution $F^{-1}$ that can be written explicitly in terms of elementary functions with the single exception of the lognormal distribution. However, for this latter case the function $C(C_G)$ can be calculated analytically.

Prior to present the results, we note that the correlation $C$ given in Eq. (\ref{cxy}) can be also expressed in terms of the standardized variables $\widetilde{x}$ and $\widetilde{y}$ (with zero mean and unit standard deviation) defined by
\begin{equation}
\widetilde{x}\equiv \frac{x-\mu}{\sigma}, \quad  \widetilde{y}\equiv \frac{y-\mu}{\sigma}
\end{equation}
with $\mu$ and $\sigma$ the corresponding mean and standard deviation of the probability density $f(x)$  (and $f(y)$). Indeed, starting from Eq. (\ref{cxy}) we can write
\begin{equation}
C=\frac{\langle x y \rangle -\mu^2}{\sigma^2}=\left\langle \left( \frac{x-\mu}{\sigma}\right) \left( \frac{y-\mu}{\sigma}\right) \right\rangle=\langle \widetilde{x} \widetilde{y}\rangle
\end{equation}

As a consequence, given any of the distributions in Table I and for any choice of the corresponding distribution parameters, the function $C(C_G)$ can be calculated simply as $\langle \widetilde{x} \widetilde{y}\rangle$. Therefore, using $\widetilde{x}$ and $\widetilde{y}$, $C(C_G)$ is given directly by Eq.(\ref{xy}) but where the inverse cumulative distribution $F^{-1}$ has to be obtained from the standardized cumulative distribution  $F(\widetilde{x})$ (and $F(\widetilde{y})$). Similarly, when using $\widetilde{x}$ and $\widetilde{y}$, the Taylor expansions given in Sec. \ref{series} have to be calculated using the inverse of $F(\widetilde{x})$ and, in addition, the expansion coefficients $b_n$ defined in Eq. (\ref{bn}) are identical
to the coefficients $a_n$ in Eq. (\ref{an}) since $\sigma^2=1$. 

In Table II we present the standardized $F(\widetilde{x})$ obtained from the cumulative distributions $F(x)$ shown in Table I using the change of
variable $x=\sigma \widetilde{x} + \mu$, with $\mu$ and $\sigma$ the mean and standard deviation of $f(x)$. Note that the standardized probability densities $f(\widetilde{x})$  can be obtained
simply as $f(\widetilde{x})=dF(\widetilde{x})/ d\widetilde{x}$.

We note that in the case of the distributions in Table I with only location and scale parameters (the cases of uniform, arcsine, logistic, Laplace and exponential distributions), the standardized
distributions $F(\widetilde{x})$ in Table II do not depend on any parameter and are therefore unique. This fact implies that for these distributions, no matter the choice of the parameters in the corresponding distributions in Table I, the function $C(C_G)$ is also unique. However, for distributions which in addition depend on a shape parameter (as the cases of symmetric Pareto, Weibull, lognormal and Pareto) the corresponding $F(\widetilde{x})$ depends also on the shape parameter and therefore is not unique but a family of distributions. Consequently, the function $C(C_G)$ is not unique either, and depend on the particular value of the shape parameter.

\begin{table}[h]
\caption{Standardized forms of the distributions shown in Table I.}
\begin{tabular}{|c|c|c|}
\hline \hline
name & support & $F(\widetilde{x})$ \\  \hline\hline
Uniform & $\left[-\sqrt{3},\sqrt{3}\right]$ & $\frac{\sqrt{3}}{6} \widetilde{x} + \frac{1}{2}$ \\ \hline
arcsine & $\left(-\sqrt{2},\sqrt{2}\right)$ & $\frac{2}{\pi} \arcsin\left( \frac{1}{2} \sqrt{\sqrt{2} \widetilde{x} +2}\right)$ \\ \hline
logistic & $\left(-\infty,\infty\right)$ & $\left[1+\exp\left(-\frac{\pi}{\sqrt{3}} \widetilde{x} \right)\right]^{-1}$ \\ \hline
Laplace & $\left(-\infty,\infty\right)$ & $\left\{  
\renewcommand{\arraystretch}{1.5}
\begin{tabular}{l}
$\frac{1}{2} \exp\left(\sqrt{2}\, \widetilde{x}\right)$ if $\widetilde{x}<0$ \\ 
$1-\frac{1}{2} \exp\left( -\sqrt{2}\, \widetilde{x} \right)$ if $\widetilde{x}\geq 0$%
\end{tabular}%
\right.  $ \\
\hline
symmetric Pareto & $\left(-\infty,\infty\right)$, $\varepsilon>2$ &  $\left\{  
\renewcommand{\arraystretch}{1.5}
\begin{tabular}{l}
$\frac{1}{2}\left(1-\sqrt{\frac{2}{(\varepsilon-1)(\varepsilon-2)}} \widetilde{x} \right)^{-\varepsilon}$ if $\widetilde{x}<0$ \\ 
$1-\frac{1}{2}\left(1+\sqrt{\frac{2}{(\varepsilon-1)(\varepsilon-2)}} \widetilde{x} \right)^{-\varepsilon}$ if $\widetilde{x}\geq 0$ \\ 
\end{tabular}%
\right.  $ 
\\ \hline
exponential & $\left[-1,\infty\right)$ & $1-\exp\left( -(\widetilde{x}+1)\right)$ \\ \hline
Weibull & 
\begin{tabular}{c}
$(-a/b,\infty)$  with \\ 
$a=\Gamma \left( \frac{1+\delta}{\delta}\right)$, $b=\sqrt{\Gamma \left( \frac{2+\delta}{\delta}\right)-a^2}$, $\delta>0$
\end{tabular}
&  $1-\exp\left( -(b \widetilde{x} +a)^{\delta} \right)$
 \\ \hline
lognormal &  $\left[ \frac{-1}{\sqrt{\exp(s^2)-1}} ,\infty\right)$, $s>0$   &  $\frac{1}{2} +\frac{1}{2}\erf\left[ \frac{\sqrt{2}}{4} s +\frac{\sqrt{2}}{2s} \ln\left(\sqrt{\exp(s^2)-1} \, \widetilde{x} +1\right) \right]$ \\ \hline
Pareto & $\left[-\sqrt{\frac{\varepsilon-2}{\varepsilon}},\infty\right)$, $\varepsilon>2$ &
$1-\left( \frac{\varepsilon-1}{\sqrt{\frac{\varepsilon}{\varepsilon-2}}\, \widetilde{x} +\varepsilon }\right)^{\varepsilon} $ \\ \hline
\end{tabular}
\end{table}

In the following, we present the results of the behavior of $C(C_G)$ for the distributions in Tables I and II. For convenience, we separate the results corresponding to 
symmetric and non-symmetric distributions.

\subsection{Symmetric distributions}

\subsubsection{Uniform distribution}

\begin{figure}[h]
\includegraphics[width=14cm]{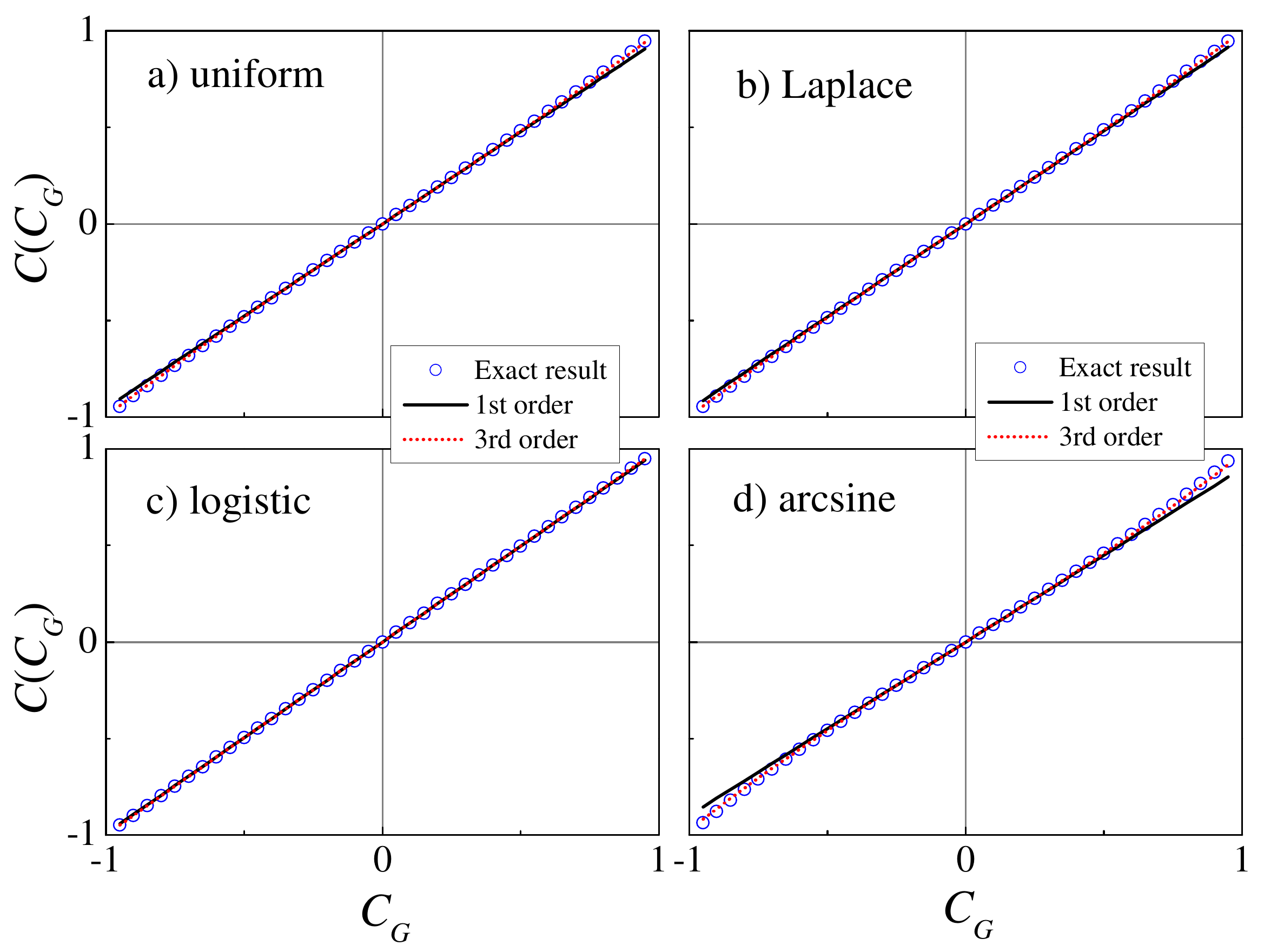}
\caption{(Color online) Correlation $C$ of the variables $x$ and $y$ as a function of the Gaussian correlation $C_G$ of the variables $x_G$ and $y_G$ when $x$ and $y$ are distributed following:
a) the uniform distribution; b) the Laplace distribution; c) the logistic distribution; and d) the arcsine distribution. In all cases, we show in circles the exact result
obtained by solving the 2D integral in Eq. (\ref{xy}), which is done analytically for the uniform distribution and numerically for the rest of the cases. We also show the results
of the expansion in Eq. (\ref{serie4}) up to first (solid lines) and third (dotted lines) orders. Note that the even order terms are null since in all cases $C(C_G)$ is an odd function.}
\label{copulas1}
\end{figure}

The uniform distribution is one of the few cases for which the function $C(C_G)$ can be obtained analytically. Since the standardized distribution $F(\widetilde{x})$ is unique in this case (see Table II),
the same happens with the function $C(C_G)$ and then it can be obtained using either $F(\widetilde{x})$  (Table II) or $F(x)$ with any choice of the parameters $a$ and $b$ (Table I).
For simplicity, we use this latter option: we start with the uniform distribution
defined in the interval $[0,1]$ so that $f(x)=1$ for $x\in[0,1]$ and $f(x)=0$ otherwise, i.e. with $a=0$ and $b=1$. This case is particularly simple since $F(\cdot)$ and $F^{-1}(\cdot)$ are
the identity function, and then the integral in Eq. (\ref{xy})
can be simplified as:
\begin{equation}
\langle x y \rangle(C_G)=\int_{-\infty}^{\infty}\int_{-\infty}^{\infty} \Phi(x_G)\, \Phi(y_G)\, \varphi_2(x_G,y_G,C_G)\, dy_G \, dx_G \label{xyu}
\end{equation}
with $\Phi(x_G)$ (and $\Phi(y_G)$) the cumulative Gaussian distribution which is given in terms of the error function (Eq. (\ref{fi})). Using the properties of the
integrals of the error function \cite{erf} and usual integration techniques, the integral in (\ref{xyu}) can be solved to obtain
\begin{equation}
\langle x y \rangle(C_G)=\frac{1}{4}+\frac{1}{2\pi}\arcsin\left(\frac{C_G}{2} \right)
\end{equation}
By introducing this result into the definition of $C$ in Eq. (\ref{cxy}), and noting that for the uniform distribution in the interval $[0,1]$ $\mu=1/2$ and $\sigma^2=1/12$
we finally get:
\begin{equation}
C(C_G)=\frac{6}{\pi}\arcsin\left(\frac{C_G}{2}\right)
\end{equation}
This function is shown in Fig \ref{copulas1}a). As expected from the general properties deduced in the previous section, we have that $C(0)=0$  and, since the uniform distribution is symmetric then $C(-C_G)=-C(C_G)$. Also, by expanding $C(C_G)$ in a Taylor series, only the odd terms are present. For the first term we have $b_1=3/\pi\simeq 0.9549$ thus indicating a strong linearity of $C(C_G)$ in this case, confirmed also by the small values of the coefficients $b_3$ and $b_5$ presented in Table III. The results of the expansion of $C(C_G)$ up to first and third order are also shown in Fig \ref{copulas1}a).

\subsubsection{Logistic, Laplace and arcsine distributions}

The three distributions, apart from being symmetric, share also another property: they lack a shape parameter. As a consequence, the corresponding $F(\widetilde{x})$ does not have any parameter and is therefore unique in the three cases (Table II), so that each distribution presents a single $C(C_G)$ function.

For the three distributions, there is no analytical solution of the integral in Eq. (\ref{xy}) which has to be solved numerically and used in Eq. (\ref{cxy}). The exact numerical results of the function $C(C_G)$ for the three cases are shown in Fig. \ref{copulas1}b), c) and d) (symbols). As expected, since the three distributions are symmetric, $C(C_G)$ is and odd function in all cases, and therefore
the corresponding Taylor expansion given by Eq. (\ref{serie4}) includes only odd terms. Indeed, we also show in Fig. \ref{copulas1} the results of the corresponding expansions up to first and third order. We note that the function $C(C_G)$ is almost linear in the three cases and the deviation from the linear behavior only occurs for extreme values of $C_G$. Specifically,
this deviation is slightly larger for the arcsine distribution but almost visually undetectable for the Laplace and  specially for the logistic case.
To quantify this almost-linear behavior, we present in Table \ref{tablacoef} the numerical results of the first three expansion coefficients $b_1$, $b_3$ and $b_5$. We recall that the coefficient
$b_j$ quantify the weight of the $j$-th term in the expansion, and then obviously the linear term is by far the one with the largest contribution: in all cases $b_1 \geq 0.9$. For the extremely
linear case of the logistic distribution, $b_1\simeq 1$ and then $C\simeq C_G$. 

\begin{table}[h]
\caption{The first three coefficients of the Taylor expansion in Eq. (\ref{serie4})
for the uniform, logistic, Laplace and arcsine distributions.} \label{tablacoef}
\begin{tabular}{c | c | c | c }
\hline
\hline
distribution & $b_1$ & $b_3$ & $b_5$  \\ \hline\hline
uniform  & $0.9549$ & $3.979\times 10^{-2}$ & $4.476\times 10^{-3}$ \\ 
logistic & $0.9919$ & $8.128\times 10^{-3}$ & $2.056\times 10^{-5}$ \\ 
Laplace  & $0.9630$ & $3.520\times 10^{-2}$ & $1.325\times 10^{-3}$ \\ 
arcsine  & $0.8995$ & $7.521\times 10^{-2}$ & $1.710\times 10^{-2}$ \\ 
\end{tabular}
\end{table}

In general, we note that the four symmetric distributions (including the uniform) present similar results, with a quite linear behavior of $C(C_G)$ since the corresponding expansion coefficient $b_1$ is the dominant one. This, together with the fact that for symmetric distributions only the odd expansion coefficients are nonzero, allows us to write
$C(C_G)\simeq b_1 C_G +O(C_G^3)$ or, in other words, the expression $C(C_G)=b_1 C_G$ is essentially correct in general for small and moderate $C_G$ values since, in addition, $b_3\ll b_1$ in all cases.
This result will prove to be important in Sec. \ref{gen-series}, where the generation of power-law correlated time series with arbitrary distribution is discussed.  
 
\subsubsection{Symmetric Pareto distribution}

For this distribution the function $C(C_G)$ has to be obtained numerically since there is no analytical solution of the integral in Eq. (\ref{xy}). The distribution is symmetric, so that $C(C_G)$ is odd.
However, $C(C_G)$ is not unique since the distribution depends on three parameters (Table I): The location parameter $\mu$ and the scale parameter $a$ (positive),  and also the shape parameter $\varepsilon$, restricted to values $\varepsilon>2$ in order to have finite variance. Then, the corresponding $F(\widetilde{x})$ 
is actually a family of distributions in terms of the shape parameter $\varepsilon$ (see Table II), which controls the power-law tail of the distribution since asymptotically $f(\widetilde{x})= dF(\widetilde{x})/d \widetilde{x} \sim |\widetilde{x}|^{-(\varepsilon+1)}$. Correspondingly, there is a family of $C(C_G)$ functions depending on the $\varepsilon$ value.

\begin{figure}
\includegraphics[width=8cm]{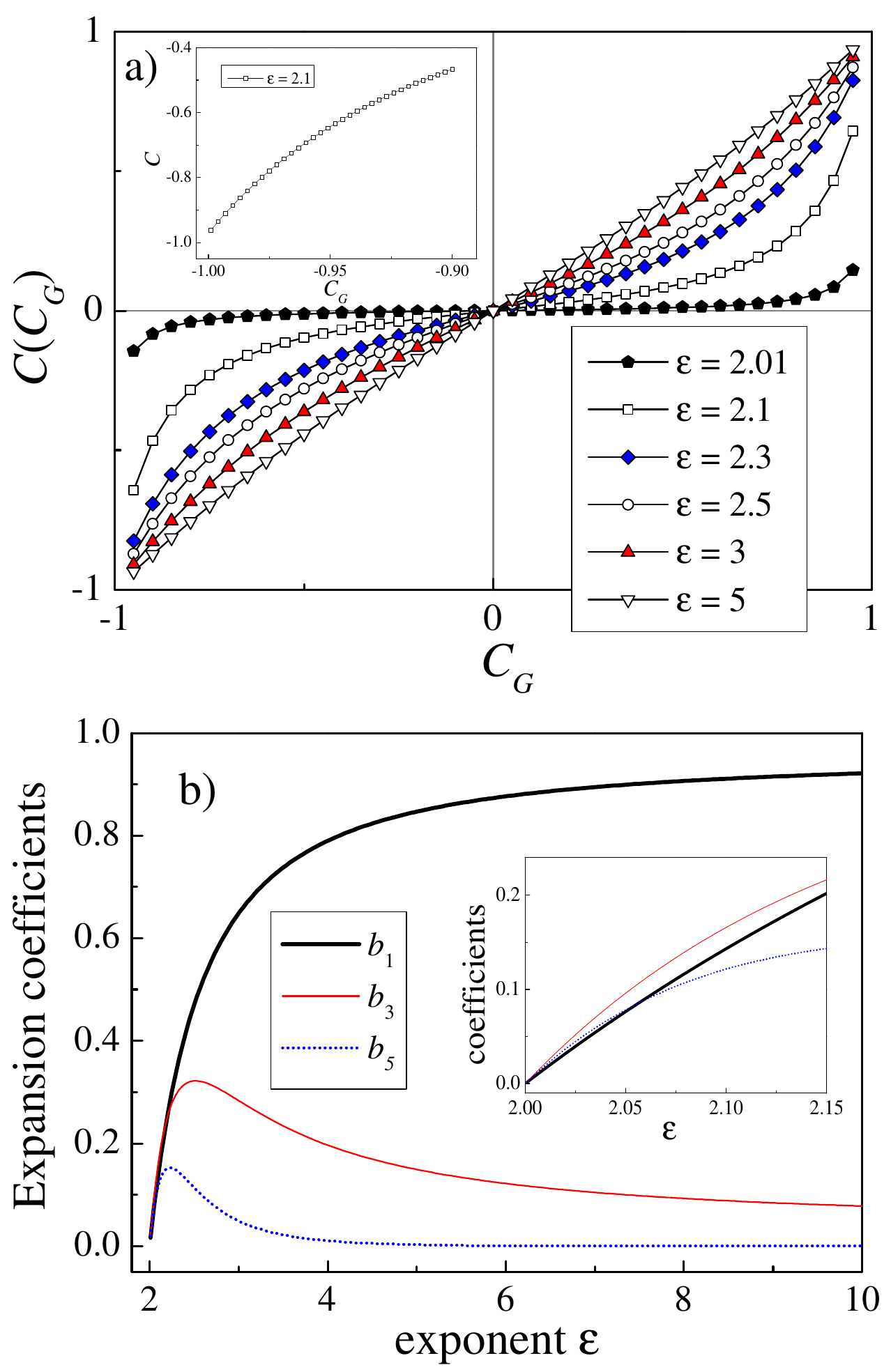}
\caption{(Color online) a) The behavior of $C(C_G)$ for different values of the exponent $\varepsilon$ controlling the power-law tails of the symmetric Pareto distribution. b) The behavior of the first
three expansion coefficients in Eq. (\ref{serie4}) as a function of $\varepsilon$. The inset shows a zoom of the region close to the limiting value $\varepsilon=2$.}  \label{figparsim}
\end{figure}

In Fig. \ref{figparsim}a) we show some $C(C_G)$ functions obtained numerically for different values of $\varepsilon$. As expected, we first note that all the functions $C(C_G)$ are odd due to the symmetry of the probability density. And second, we also find that the linearity of the $C(C_G)$ function decreases as the $\varepsilon$ value becomes smaller: while for large $\varepsilon$ values (fast-decaying power-law tail) $C(C_G)$ behaves quite linearly, as $\varepsilon$ decreases (longer power-law tail) and approaches the limiting value $\varepsilon=2$, the function $C(C_G)$ becomes smaller and more nonlinear, flattens and eventually tends to 0 as $\varepsilon \rightarrow 2$. This effect can be quantified by calculating the expansion coefficients in Eq. (\ref{serie4}), which account for the 
weights of the successive expansion terms. In Fig. \ref{figparsim}b) we plot the first non-zero expansion coefficients ($b_1$, $b_3$ and $b_5$) obtained numerically from Eq. (\ref{bn}). We note that
for large $\varepsilon$ values, $b_1$ is the largest coefficient and close to one, confirming the strong linearity of $C(C_G)$ in this $\varepsilon$ range. However, for smaller $\varepsilon$ values, we observe that all the expansions coefficients tend to zero as $\varepsilon \rightarrow 2$ (see also the inset in Fig. \ref{figparsim}b)), in agreement with the flattening of $C(C_G)$ around $C=0$.

These results indicate that when transforming Gaussian variables $x_G$ and $y_G$ with correlation $C_G$ into the variables $x$ and $y$ following the symmetric Pareto distribution, the correlation $C$ of $x$ and $y$ is smaller as the power-law tail of the distribution, controlled by $\varepsilon$, becomes longer. Eventually, $x$ and $y$ will be uncorrelated in the limit $\varepsilon\rightarrow 2$. The implication of this property in the generation of time series following the symmetric Pareto distribution will be discussed in Sec. \ref{gen-series}.

\subsection{Non-symmetric distributions}

\subsubsection{lognormal distribution}

We start with the case of lognormal distribution because it is the only one (together with the uniform distribution discussed above) for which the function $C(C_G)$ can be obtained analytically.
In this case, noting that  $F^{-1}(y)=\exp \left( m +\sqrt{2} s \erf^{-1}(2y-1)\right)$ (Table I), and that $\Phi(x_G)$ is given in Eq. (\ref{fi}) we obtain that
\[
F^{-1}(\Phi(x_G))=\exp(m+ s x_G)
\]
and similarly for $F^{-1}(\Phi(y_G))$. Then, the integral in Eq. (\ref{xy}) is given by
\begin{equation}
\langle x y \rangle(C_G)=\int_{-\infty}^{\infty}\int_{-\infty}^{\infty} \exp(m+ s x_G) \, \exp(m+ s y_G)\, \varphi_2(x_G,y_G,C_G)\, dy_G \, dx_G \label{xylogn}
\end{equation}
which can be evaluated to obtain
\begin{equation}
\langle x y \rangle(C_G)=\exp[(C_G+1)s^2+2m] \label{intlog}
\end{equation}

\begin{figure}[ht]
\includegraphics[width=8cm]{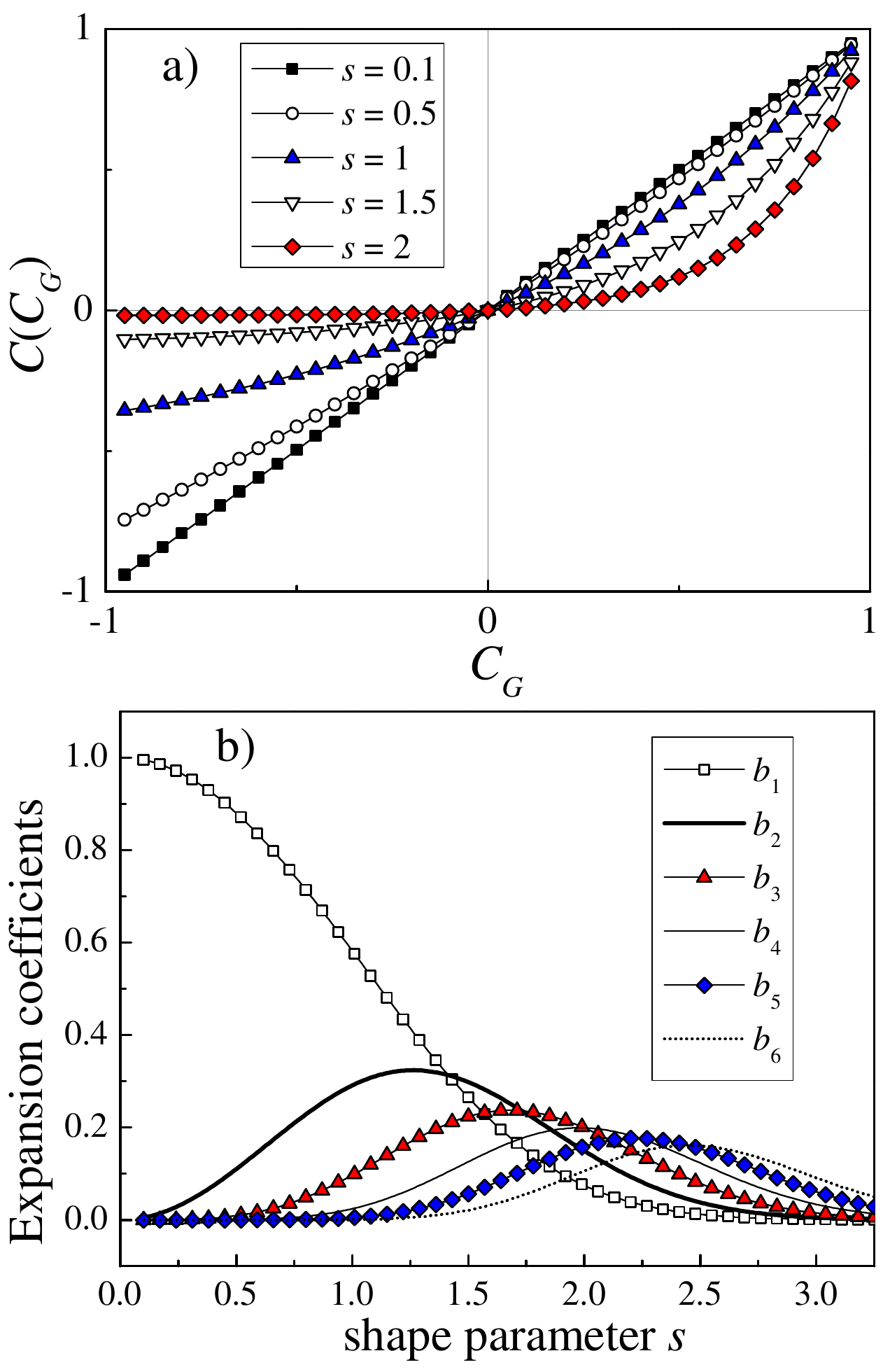}
\caption{(Color online) a) Behavior of the correlation $C$ of two lognormally distributed variables as a function of the correlations $C_G$ (Eq. (\ref{res-logn})) of the two Gaussian variables
from where the lognormal variables have been obtained. The different curves correspond to different values of the shape parameter $s$ of the lognormal distribution.
b) Behavior of the first six expansion coefficients $b_n$  in Eq. (\ref{serie2}) as a function of the lognormal shape parameter $s$. For the lognormal case, $b_n$ 
can be obtained analytically (see. Eq. (\ref{blog})). }\label{fig-logn}
\end{figure}

Since for the lognormal distribution the mean $\mu$ and the variance $\sigma^2$ are given respectively by $\mu=\exp(m+s^2/2)$ and $\sigma^2=[\exp(s^2)-1] \exp(2m+s^2)$,
we can insert these values and the result for $\langle x y \rangle(C_G)$ in Eq. (\ref{intlog}) into Eq. (\ref{cxy}) to obtain finally 
\begin{equation}
C(C_G)=\frac{\exp(s^2 C_G) -1}{\exp(s^2)-1} \label{res-logn}
\end{equation}

As expected from the standardized version of the lognormal distribution in Table II, the function $C(C_G)$ is not unique but a family of functions controlled by the
shape parameter of the distribution, $s$, which is restricted to positive values. We show in Fig. \ref{fig-logn}a) the function $C(C_G)$ for several
values of $s$. First, we note, as expected, that in this case the function $C(C_G)$ is not odd, but the sign of the correlations is preserved, i.e., for positive $C_G$ values, $C(C_G)>0$ and $C(-C_G)<0$. The $C_{\min}$ value increases (decreases in absolute value) with the parameter $s$, which controls the tail of the lognormal distribution, longer for larger $s$. Indeed, for moderately large $s$ values it is almost impossible to get anticorrelated lognormal variables since $C(C_G)$ is practically zero for $-1<C_G<0$ (see the case $s=2$ in Fig. \ref{fig-logn}a)), while the behavior for $0<C_G<1$ is substantially different. The $C_{\min}$ behavior is systematically studied in Sec. \ref{feasibility}.
And second, we also note that $C(C_G)$ becomes more nonlinear as the shape parameter $s$ increases. The degree of nonlinearity can be quantified again using the expansion coefficients $b_n$
in Eq. (\ref{bn}) which in this case can be obtained analytically by expanding in a Taylor series Eq. (\ref{res-logn}): 
\begin{equation}
b_n=\frac{s^{2n}}{(\exp(s^2)-1) n!} \label{blog}
\end{equation}

The behavior of the first six expansion coefficients as a function of $s$ is depicted in Fig. \ref{fig-logn}b), and confirms the observed behavior of $C(C_G)$: while for small $s$ the linear behavior
in $C(C_G)$ dominates, for increasing $s$ the linear coefficient $b_1$ tends to zero and, depending on the $s$ range, a different expansion coefficient is the dominant one. These results imply that
the validity of the linear approximation $C(C_G)\simeq b_1 C_G$ depends on the $s$ value: while for small $s$ values the linear approximation is essentially correct for small and moderately large
$|C_G|$ values, for large $s$ the linear approximation will be correct only for very small $|C_G|$ values since $b_1$ will be the smallest coefficient in this range, and then very small $|C_G|$ values
are required to neglect higher order expansion terms. This fact will affect the possible generation of power-law correlated, lognormally distributed time series (see Sec. \ref{gen-series}).

\subsubsection{Exponential distribution}

\begin{figure}[h]
\includegraphics[width=8cm]{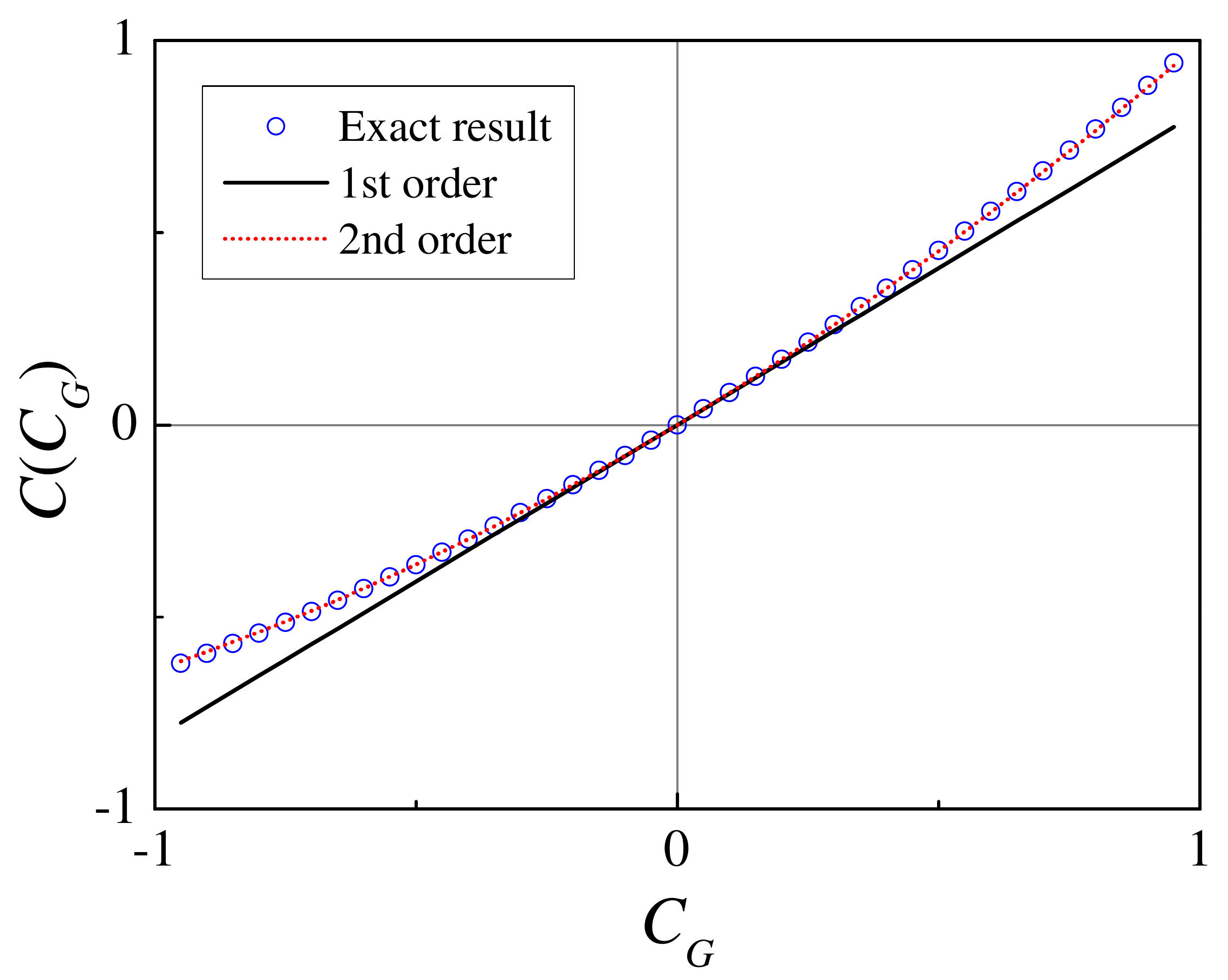}
\caption{(Color online) Behavior of $C(C_G)$ for the exponential distribution. We show in symbols the exact result obtained by solving numerically the integral in Eq. (\ref{xy}). We also
show the Taylor expansion of $C(C_G)$ according to Eq. (\ref{serie2}) up to first (solid line) and second (dotted line) orders.}\label{fig-expo}
\end{figure}

The exponential distribution lacks a shape parameter, and then there is a single standardized exponential distribution so that the function $C(C_G)$ is unique (Table II). However,
there is no analytical solution of the integral in Eq. (\ref{xy}) in this case, which has to be calculated numerically. The exact numerical result of the $C(C_G)$ function for
the exponential distribution is shown in Fig. \ref{fig-expo}.

Again, as the exponential distribution is not symmetric, the function $C(C_G)$ is not odd either. However, $C(C_G)$ is fairly linear, specially for intermediate and small $|C_G|$ values.
This linearity can be quantified by calculating the corresponding expansion coefficients
$b_n$ in Eq. (\ref{bn}). The first 4 coefficients result to be: $b_1=0.8158$, $b_2=0.1774$,
$b_3=6.684\times 10^{-3}$ and $b_4=1.343 \times 10^{-4}$. Indeed, in Fig. \ref{fig-expo} we also
show the expansions of $C(C_G)$  according to Eq. (\ref{serie2}) up to first and second orders. In this case, the second order expansion is very precise in the whole $C_G$ range, and the first order suffices for  $|C_G|<0.1$. This property implies that it would be possible to generate power-law correlated and exponentially distributed time series (see Sec. \ref{gen-series}).

\subsubsection{Weibull and Pareto distributions}

\begin{figure}[h]
	\includegraphics[width=16cm]{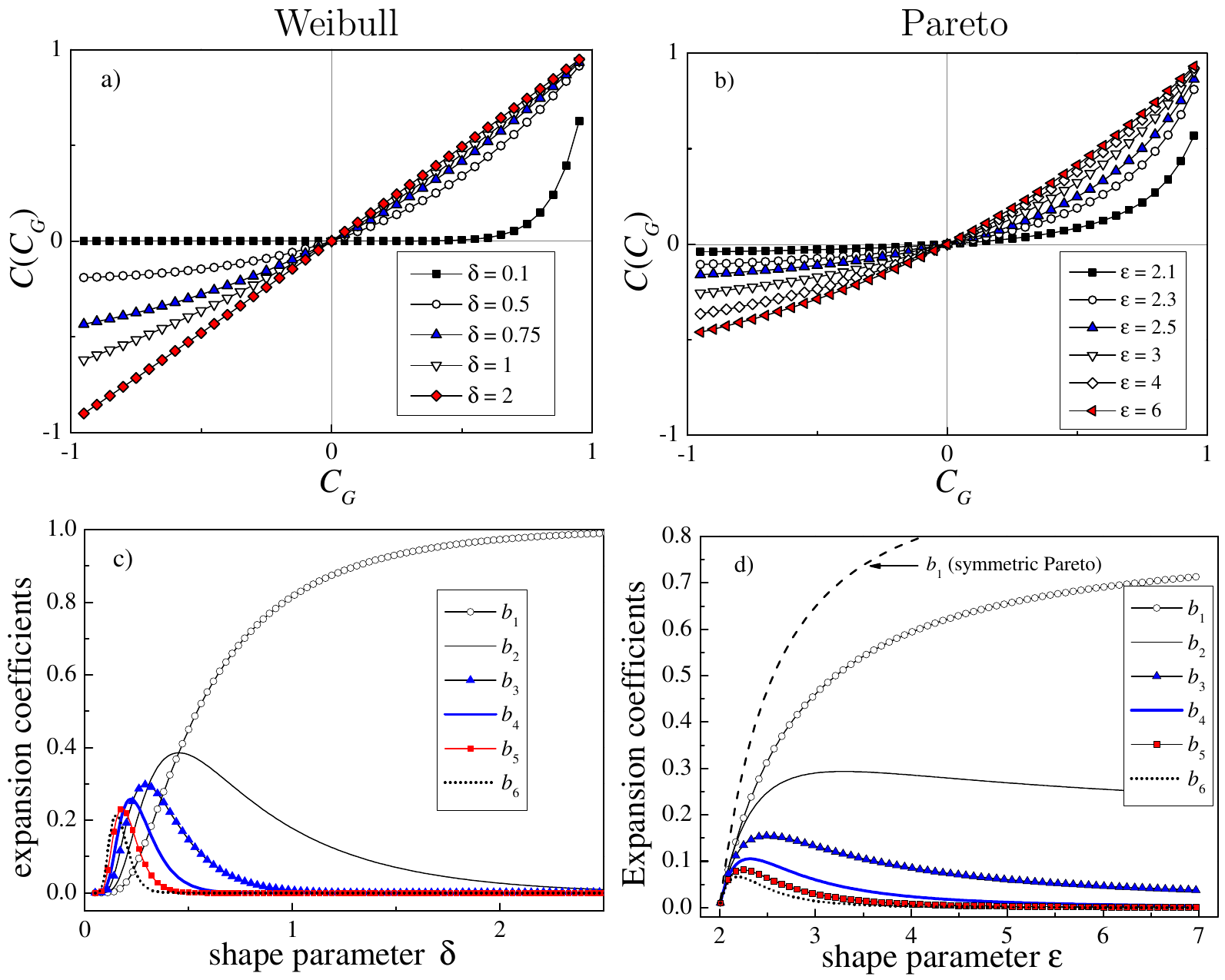}
	\caption{(Color online) Behavior of the correlation $C$ of two variables following a Weibull distribution (panel a)) and a Pareto distribution (panel b)) as a function of the correlations $C_G$ of the two Gaussian variables from where the Weibull and the Pareto variables have been obtained using Eq. (\ref{tr}). The different curves correspond to different values of the corresponding shape parameters $\delta$ and $\varepsilon$. Behavior of the first six expansion coefficients $b_n$  in Eq. (\ref{serie2}) as a function of the shape parameter $\delta$ of the Weibull distribution (panel c)) and of the shape parameter
$\varepsilon$ of the Pareto distribution (panel d)).}\label{fig-wei-par}
\end{figure}

We present these two distributions together because the corresponding $C(C_G)$ functions present similar properties. As can be seen in Table II, the standardized forms of the Weibull and
Pareto distributions have shape parameters, $\delta$ and $\varepsilon$ respectively, which control the tail behavior. For the Weibull case, the parameter $\delta$ must be
positive, $\delta>0$. In the range $\delta>1$, the tail decays faster than exponentially, and the larger $\delta$, the faster the decay; the case 
$\delta=1$ corresponds to the exponential distribution, that we have studied above; and the case $\delta<1$ corresponds to a heavy tail distribution with a decay slower than exponential (stretched-exponential form),  and the smaller $\delta$ the longer the tail of the distribution. For the Pareto case, the exponent $\varepsilon>2$ controls
the power-law tail of the distribution, with a probability density $f(x)$ with asymptotic behavior $f(x) \sim x^{-(\varepsilon+1)}$. For both the Weibull and Pareto distributions, the existence
of shape parameters implies that the function $C(C_G)$ is not unique but a family of functions controlled by $\delta$ and $\varepsilon$, respectively. In the two cases, the integral in Eq. (\ref{xy}) does not admit an analytical solution, and has to be solved numerically. In Fig. \ref{fig-wei-par}a) and \ref{fig-wei-par}b) we show several examples of $C(C_G)$ functions obtained numerically
for different values of the Weibull shape parameter $\delta$ and the Pareto shape parameter $\varepsilon$, respectively. 

Since both distributions are not symmetric, we first observe, as expected, that $C(C_G)$ is not odd, but the sign of the correlations is preserved for both distributions. We also observe that the corresponding $C_{\min}$ increases (decreases in absolute value) as $\delta$ and $\varepsilon$ decrease and the tail of the distributions becomes heavier, so that it will be difficult to obtain
Weibull and Pareto anticorrelated variables (See Sec. \ref{feasibility}). In the limits $\delta\rightarrow 0$ and  $\varepsilon\rightarrow 2$, $C(C_G)$ tends to zero in the whole $C_G$ range. Similarly, the degree of nonlinearity of $C(C_G)$ is also controlled by $\delta$ and $\varepsilon$: while for large $\delta$ and $\varepsilon$ values (fast-decaying tails) $C(C_G)$ is more linear, as $\delta$ and $\varepsilon$ decrease and tend to the respective limits 0 and 2, the function $C(C_G)$ becomes strongly nonlinear. As in previous cases, the nonlinear behavior can be quantified by calculating the expansion coefficients $b_n$ of the function $C(C_G)$ defined in Eq. (\ref{bn}).  In Fig. \ref{fig-wei-par}c) and Fig. \ref{fig-wei-par}d)  we plot the first six expansion coefficients as a function of the shape parameters $\delta$ and $\varepsilon$, respectively.

We obtain that for large $\delta$ and $\varepsilon$ values the linear term is by far the most important, even with $b_1$ values close to 1 indicating an almost perfect linear behaviour of $C(C_G)$ (see the
case $\delta=2$ in Fig. \ref{fig-wei-par}a)). As $\delta$ and $\varepsilon$ decrease, $b_1$ becomes smaller indicating a loss of linearity, and higher order expansion coefficients can be important. These results imply that the first-order approximation $C(C_G) \simeq b_1 C_G$ is essentially correct for small and moderately large $|C_G|$ values if the shape parameters $\delta$ and $\varepsilon$ are large. However, for small $\delta$ and $\varepsilon$ values, and especially for $\delta$ values close to zero and $\varepsilon$ values close to 2, the approximation will be valid only for very small $|C_G|$ values. The implications of this fact when generating power-law correlated times series following Weibull and Pareto distributions are discussed in the next section.

\subsection{Feasible correlations for non-symmetric distributions} \label{feasibility}

We have shown above that the function $C(C_G)$ is increasing, maps positive (negative) $C_G$ values into positive (negative) $C$ values, with $C(0)=0$ and $C(1)=1$. For symmetric distributions, in addition, $C(C_G)$ is odd, so that $C(-1)=-1$ and then the final non-Gaussian variables $x$ and $y$ can be correlated with any value in the interval $(-1,1)$. However,
for non-symmetric distributions, $C(-1)=C_{\min}$ with $-1<C_{\min}<0$. This means that for this kind of distributions, the $C$ values in the interval $(-1,C_{\min})$ are not reachable no matter
the original $C_G$ value, and therefore the range of \textsl{feasible} correlations corresponds to the interval $(C_{\min},1)$.

\begin{figure}[h]
	\includegraphics[width=8cm]{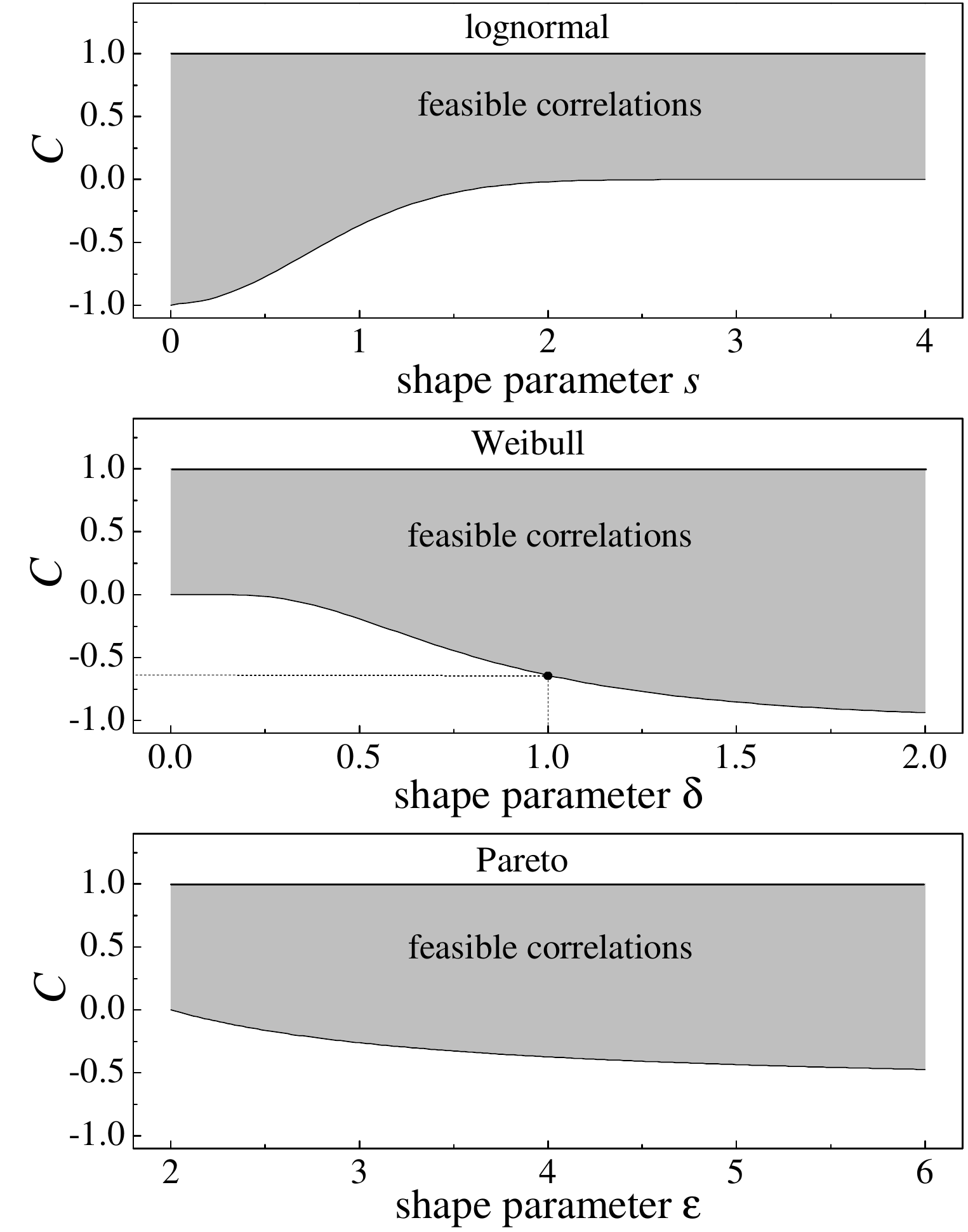}
	\caption{Interval $(C_{\min},1)$ of feasible correlations for $x$ and $y$ variables obtained via the transformation (\ref{tr}) when their marginal distribution is, from top to bottom,
	lognormal, Weibull and Pareto. In all panels the botton curve represents the $C_{\min}$ value as a function of the corresponding shape
	parameter, so that the shaded areas show the feasible correlations in each case. The exponential distribution corresponds to the case $\delta=1$ in the Weibull case, for which
	$C_{\min}\simeq -0.64$ and is shown with a solid circle.} \label{fig-cmin} 
\end{figure}

In general, the particular value of $C_{\min}$ depends on the final marginal non-symmetric distribution considered, and for a given distribution $C_{\min}$ can be calculated by evaluating the integral (\ref{xy}) using $C_G$ values close to $-1$. A non-symmetric distribution with shape parameter corresponds actually to a family of distributions, so that $C_{\min}$ is also a function of the particular value of the shape parameter. We have determined the interval of feasible correlations $(C_{\min},1)$ for the non-symmetric distributions of Table I, which are shown in Fig. \ref{fig-cmin}. 
Since the lognormal, Weibull and Pareto cases have a shape parameter, the bottom curve in the three panels shows the value of $C_{\min}$ as a function
of the respective shape parameter, so that the feasible correlations correspond to the shaded areas. The exponential distribution is a particular case of the Weibull distribution for 
$\delta=1$, for which $C_{\min} \simeq 0.64$ and is indicated in the central panel of Fig. \ref{fig-cmin}) with a solid circle.

In general, we observe that the longer the tail of the non-symmetric distribution considered (controlled by its shape parameter), the larger the $C_{\min}$ value (smaller in absolute value) and the shorter
the interval of feasible correlations. In the extreme cases of very heavy tails, $C_{\min}$ can be practically 0, so that it is almost impossible to obtain negative $C$ values, or in other words, is it almost impossible to get anticorrelated $x$ and $y$ variables via the transformation (\ref{tr}) when the final marginal distribution is very long-tailed. For the lognormal and Weibull cases, $C_{\min}\simeq 0$ even for shape parameter values not even close to their limiting values ($s\rightarrow \infty$ and $\delta=0$ respectively), so that we see a practically flat $C_{\min}$ curve in both cases when approaching the limiting values. For the Pareto case, we get $C_{\min}\rightarrow 0$ for $\varepsilon \rightarrow 2$ but the  $C_{\min}$ curve is not flat but decreasing when $\varepsilon$ increases.

\section{Application to time series}

We have obtained the results of Secs. II and III by transforming two correlated Gaussian variables $x_G$ and $y_G$ into two variables $x$ and $y$ with the same arbitrary marginal distribution. 
These results can be naturally extended to the transformation of Gaussian correlated time series into time series with arbitrary marginal distribution, as we stated in the Introduction.
Let us consider a $\sim {\cal N}(0,1)$ Gaussian time series $\{z_{G,i}\}$, $i=1,2,\ldots,N$, with autocorrelation function $C_G(\ell)\equiv \langle z_{G,i} z_{G,i+\ell}\rangle$
than can be calculated for any value of the lag $\ell$. We can transform the Gaussian time series $\{z_{G,i}\}$ into a time series $\{z_{i}\}$ with arbitrary marginal distribution using (\ref{transforma1}). The autocorrelation function $C(\ell)\equiv (\langle z_i z_{i+\ell}\rangle - \mu^2)/\sigma^2$ of $\{z_{i}\}$ is then determined by the behavior of the $C(C_G)$ function. Indeed, simply by replacing back 
in Eq. (\ref{xy}) $x_G$ and $y_G$ by $z_{G,i}$ and $z_{G,i+\ell}$ respectively, $x$ and $y$ by $z_i$ and $z_{i+\ell}$, and also $C_G$ by $C_G(\ell)$ we obtain automatically that
\begin{equation}
C(\ell)=C(C_G(\ell)) \label{result}
\end{equation}
i.e., the autocorrelation function of the final time series is determined by the $C(C_G)$ function (depending only on the final marginal distribution) and
the Gaussian autocorrelation function $C_G(\ell)$. We remark that this last result is correct since the Gaussian time series
posseses a well-defined autocorrelation function, and the $C(C_G)$ function can be obtained
using the integral (\ref{xy}) for any final marginal distribution and for any value of $C_G\in(-1,1)$, and therefore there are no feasibility problems when creating the final non-Gaussian time series
since we are obtaining $C(\ell)$ for the corresponding marginal distribution, and not imposing it. Note that feasibility problems can appear when imposing in a time series a marginal distribution and also an specific autocorrelation function, and both properties may not be compatible \cite{cario}. For example, for non-symmetric distributions with long tails, negative $C(\ell)$ values are likely non-feasible
(see Sec. \ref{feasibility}). The validity of Eq. (\ref{result}) for time series, inherited from Eq. (\ref{xy}) for pairs of variables, has been previously discussed for example in \cite{kugiumtzis,kugi02}.

To illustrate the applicability of our results to time series, we consider two examples of Gaussian time series with well-defined autocorrelation functions $C_G(\ell)$, which are then transformed to have two different marginal distributions. The first Gaussian time series  we consider are autoregressive processes of order 1, AR(1), defined as
\begin{equation}
z_{G,i}=\varphi z_{G,i-1} + \eta_{i}
\end{equation}
where $\{\eta_i\}$ a Gaussian white noise such that $\eta_i \sim {\cal{N}} (0,1)$ and $\varphi \in(-1,1)$ is a constant. AR(1) processses are Gaussian, and the corresponding autocorrelation function $C_G(\ell)$ is given by:
\begin{equation} 
C_G(\ell) = \varphi^{\ell}
\end{equation}
equivalent to an exponentially decreasing function, of alternate sign for $\varphi<0$.
The second example of Gaussian time series are the outputs of the Fourier Filtering method,
described in more detail below (see Sec. \ref{gen-series}), which present a power-law autocorrelation function with
exponent controlled by the Hurst exponent $H \in (0,1)$ \cite{Hurst} (see Eq. (\ref{cor-fgn})). 

\begin{figure}[h]
\includegraphics[width=8cm]{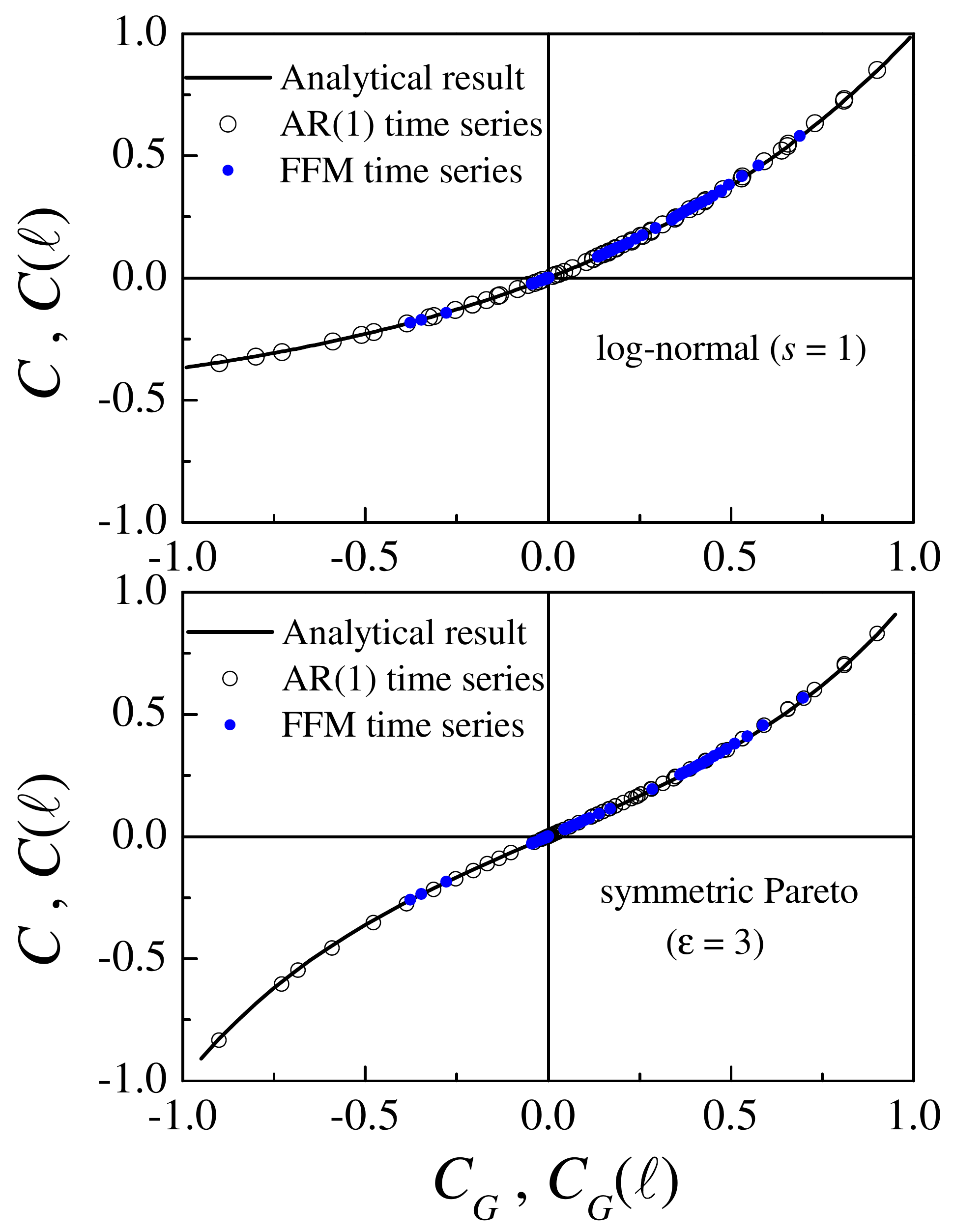}
\caption{Top panel: the theoretical $C(C_G)$ function for a lognormal final marginal distribution (solid line). We also generate several Gaussian FFM and AR(1) time series with different
$H$ and $\varphi$ values, calculate the corresponding $C_G(\ell)$ and then transform the time series to have the considered lognormal marginal distribution and calculate $C(\ell)$. The symbols correspond
to plot $C(\ell)$ vs. $C_G(\ell)$ for both kind of time series, and fall perfectly on top of the theoretical curve. Bottom panel: the same as in the top panel, but for a symmetric Pareto
final marginal distribution.} \label{compara}
\end{figure}

In Fig. \ref{compara} we consider two different final marginal distributions, lognormal (top panel) and symmetric Pareto (bottom panel). First, we represent as solid lines the theoretical $C(C_G)$
functions obtained as explained in Secs. II and III. Then, we generate several AR(1) and FFM Gaussian time series with different $\varphi$ and $H$ parameters. For each time series, we start calculating
the autocorrelation function $C_G(\ell)$, then the time series is transformed to have the final marginal distribution considered using Eq. (\ref{transforma1}), and finally we obtain the autocorrelation function $C(\ell)$ and represent $C(\ell)$ vs. $C_G(\ell)$, as shown in symbols in Fig. \ref{compara}. We note that the symbols fall perfectly on top of the theoretical $C(C_G)$ functions, independently of
the final marginal distribution or the values of $\varphi$ and $H$, showing the validity of Eq. (\ref{result}).  
 
We also remark that although the results in Secs. II and III for the $C(C_G)$ function have been obtained for final marginal distributions with known analytical expressions, the same technique can be applied to experimental time series for which the marginal distribution is not known analytically. Indeed, it is enough to determine numerically the cumulative distribution $F(x)$ and its inverse $F^{-1}(y)$, and use it in the numerical solution of the integral in Eq. (\ref{xy}) to obtain how the correlations change when transforming correlated Gaussian variables into variables with the same marginal distribution of the experimental data. Indeed, we use this approach in one of the applications addressed below.

\subsection{Application I: Generation of power-law correlated time series with arbitrary distribution} \label{gen-series}

Probably, fractional Gaussian noises (fGns) \cite{Beran} are the reference for stochastic Gaussian time series with power-law autocorrelation functions. The autocorrelation function
of a fGn is given by
\begin{equation}
C_G(\ell)=\frac{(\ell-1)^{2 H}-2\ell^{2H}+(\ell+1)^{2H}}{2} \label{fgn-exacto}
\end{equation}
where $H\in(0,1)$ is the Hurst exponent \cite{Hurst}. The power-law nature of $C_G(\ell)$ arises in the limit of large $\ell$ where we have
\begin{equation} 
C_G(\ell)\simeq \frac{H(2H-1)}{\ell^{2-2H}}\propto \frac{{\rm sign}(H)}{\ell^{2-2H}} \label{cor-fgn}
\end{equation}
The case $H=1/2$ corresponds to absence of correlations (white noise); the case $1/2<H<1$ corresponds to positive correlations,
which decay slower with $\ell$ for larger $H$ values; and the case $0<H<1/2$ corresponds to negative correlations,
which decay faster (in absolute value) as $H$ becomes smaller.

Likely, the algorithm most widely used to generate Gaussian power-law correlated time series of fGn type in different contexts is the Fourier Filtering Method (FFM) 
\cite{FFM1,FFM2,Fidelis,Conchita,Yosi_volatility,cor-size,manolo,carpena_dfa,bernaola,uso_ffm_1,Hu2001,ChenPRE2005,escalas,super}. Although there are different approaches to implement FFM, probably the simplest is the following: 1) Given a time series size $N$, consider a power spectrum as
\begin{equation}
S(f_j) \propto f_j^{2H-1} \,\, {\rm with} \,\, f_j=\frac{j}{N}, j=1,2,\ldots,N/2 \label{power}
\end{equation}
with $H\in(0,1)$ the input Hurst exponent. 2) Construct a Fourier transform such that $\Re(F(f_j))=S(f_j)^{1/2} \cos(\phi_j)$ and $\Im(F(f_j))=S(f_j)^{1/2} \sin(\phi_j)$ with $\phi_j$ a random phase
uniformly distributed in the interval $[0,2\pi]$. 3) Fourier-transform back $F(f_j)$ into real space to obtain the Gaussian time series  $\{z_{G,i}\}$, $i=1,2,\ldots,N$.
By construction, the power spectrum of $\{z_{G,i}\}$ is given by (\ref{power}), and then, via the Wiener-Khinchin theorem, the autocorrelation function $C_G(\ell)$ of $\{z_{G,i}\}$
is power-law behaved as in Eq. (\ref{cor-fgn}) with well-defined Hurst exponent $H$. In addition to power-law correlated, Gaussian and stationary, $\{z_{G,i}\}$ is also
purely linear, since the Fourier phases are random. Without loss of generality, we can normalize $\{z_{G,i}\}$ to have zero mean and unit standard deviation,
$z_{G,i} \sim {\cal N}(0,1)$ $\forall i$, and then with probability density $\varphi(z_G)$ and cumulative
distribution $\Phi(z_G)$ as the ones given in Eq. (\ref{fi}).

\begin{figure}
\includegraphics[width=12cm]{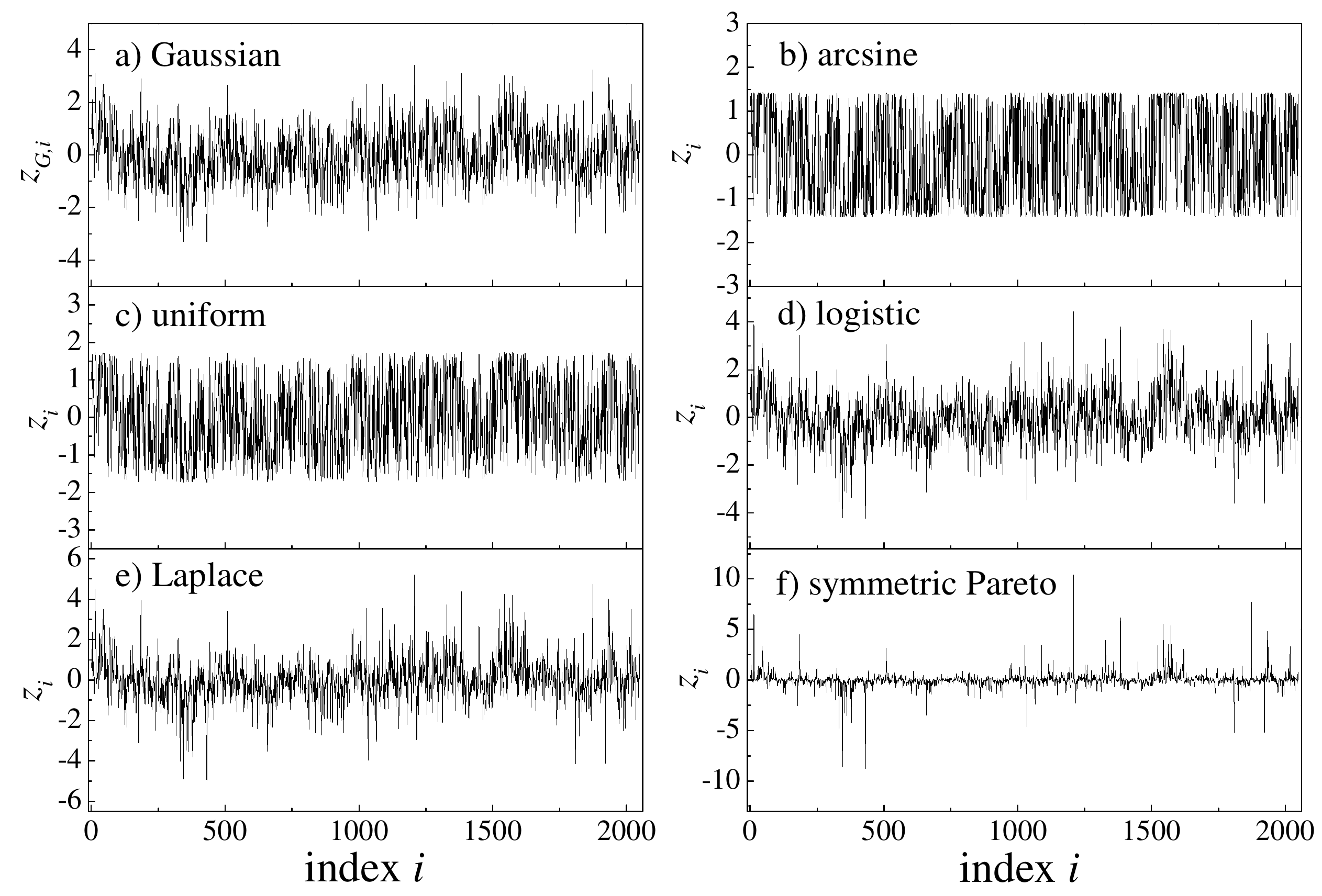}
\caption{a) Gaussian time series $\{z_{G,i}\}$ of zero mean and unit standard deviation generated using FFM with $H=0.8$ and size $N=2^{11}$. The rest of the panels
show time series $\{z_i\}$ following the symmetric distributions in Table I which are obtained by transforming $\{z_{G,i}\}$ using Eq. (\ref{transforma1}). The different
panels correspond to: b) Arcsine distribution; c) Uniform distribution; d) Logistic distribution; e) Laplace distribution; f) Symmetric Pareto distribution with shape parameter $\varepsilon=2.3$. In all cases, the distributions have zero mean and unit standard deviation.} \label{ser-sim}
\end{figure}

\begin{figure}
\includegraphics[width=12cm]{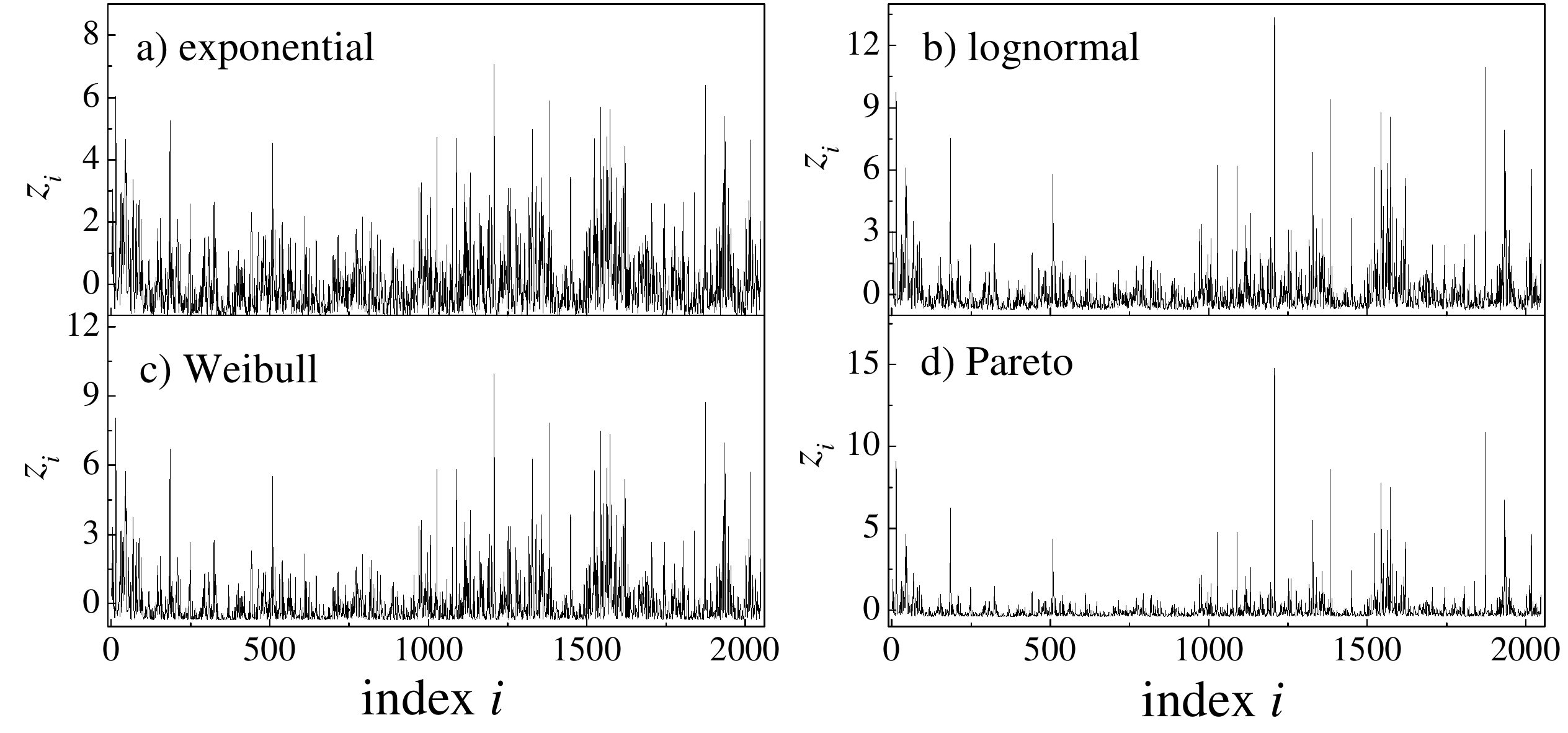} 
\caption{Time series $\{z_{i}\}$ following the non-symmetric distributions in Table II obtained by transforming the Gaussian time
series $\{z_{G,i}\}$ shown in Fig. \ref{ser-sim}a) using Eq. (\ref{transforma1}). The panels correspond to: a) Exponential distribution; b) lognormal distribution
with shape parameter $s=1$; c) Weibull distribution with shape parameter $\delta=0.7$; d) Pareto distribution with shape parameter $\varepsilon=2.3$. In all cases,
the distributions have zero mean and unit standard deviation.} \label{ser-nosim}
\end{figure}

We suggest here to use FFM as the initial step of the algorithm able to generate power-law correlated time series $\{z_i\}$ with arbitrary distribution and controlled $H$.
Once the time series $\{z_{G,i}\}$ is generated with FFM, we propose to use Eq. (\ref{transforma1}) to transform $\{z_{G,i}\}$ into time series $\{z_i\}$ with arbitrary marginal
distribution. As an example, in Figs. \ref{ser-sim} and \ref{ser-nosim} we show several time series $\{z_i\}$ obtained via Eq. (\ref{transforma1}) from a Gaussian power-law correlated time series $\{z_{G,i}\}$ (shown in Fig. \ref{ser-sim}a)) generated with FFM. The final marginal distributions of $\{z_i\}$ correspond to the distributions in Table I, with the symmetric cases shown in
Fig. \ref{ser-sim}, and the non-symmetric ones in Fig. \ref{ser-nosim}.

Since the marginal distribution of the final time series ${\{z_i\}}$ is controlled via Eq. (\ref{transforma1}), the important question is whether the $\{z_i\}$ series are also power-law correlated with well defined Hurst exponent $H$ or not. We have shown above (Eq. (\ref{result}) that $C(\ell)=C(C_G(\ell))$. Therefore $C(\ell)$ will show a power-law behavior with the same Hurst exponent as $C_G(\ell)$ when $C(C_G)$ behaves linearly. According to the expansion in Eq. (\ref{serie2}), this happens whenever the first term in the expansion is the
dominant one. In such case we can write 
\begin{equation}
C(\ell) \simeq b_1 C_G(\ell) \label{linearC}
\end{equation}
We note that, theoretically speaking, this linear approximation would be always correct for sufficiently small $C_G(\ell)$, where higher powers of $C_G(\ell)$ can be neglected in the expansion. As $C_G(\ell)$ is a decaying power-law (\ref{cor-fgn}), this means that the linear approximation will ultimately work for large enough $\ell$ and then, in the limit of large $\ell$, $C(\ell)$ will tend asymptotically to a power-law of the type written in Eq. (\ref{linearC}) with the same Hurst exponent $H$ as $C_G(\ell)$, no matter the distribution of  the final time series $\{z_i\}$.  

However, the asymptotic validity of the linear approximation (\ref{linearC}) does not suffice in practical purposes to generate time series with observable power-law correlations. The reason is two-fold, since it depends on the length $N$ of the time series $\{z_{G,i}\}$ and $\{z_i\}$, and the $b_1$ value of the $\{z_i\}$ marginal distribution. Note that for a FFM-generated $\{z_{G,i}\}$ time series
of length $N$, the expected noise level of the autocorrelation values is about $2/\sqrt{N}$ \cite{Beran}, and then values below this level are not significant.
Indeed, a more precise value for the noise level of $C_G(\ell)$ is $2/\sqrt{N-\ell}$, since only $N-\ell$ samples can be used to estimate $C_G(\ell)$. Therefore, 
the maximum value of the lag $\ell$ up to which there is observable and significant power-law correlated behavior  in the Gaussian time series, $\ell_{G,\max}$,  can be estimated by equating the autocorrelation function $C_G(\ell)$ (\ref{fgn-exacto}) at $\ell=\ell_{G,\max}$ and the corresponding noise level
\begin{equation}
C_G(\ell_{G,\max})=\frac{2}{\sqrt{N-\ell_{G,\max}}}
\end{equation}
and solving numerically for $\ell_{G,\max}$. The solution, obviously, depends on $N$ and $H$, and in general increases with $N$ and $H$. 

Similarly, we can estimate the maximum lag, $\ell_{\max}$, up to which the linear approximation to $C(\ell)$ (Eq. (\ref{linearC}) presents significant values by solving the equation
\begin{equation}
b_1 \left( \frac{(\ell_{\max}-1)^{2 H}-2\ell_{\max}^{2H}+(\ell_{\max}+1)^{2H}}{2}\right)=\frac{2}{\sqrt{N-\ell_{\max}}} \label{criterio}
\end{equation}
where we write the explicit expression of $C_G(\ell_{\max})$. The solution of this latter equation depends on $N$ and $H$, and also on the marginal distribution of $\{z_i\}$ via its $b_1$ value. 
In general, $\ell_{\max}$ increases with $N$, $H$ and $b_1$, and since $b_1<1$, $\ell_{\max}<\ell_{G,\max}$. 

Indeed, given $N$, $H$ and a final marginal distribution
for $\{z_i\}$ with a particular $b_1$ value, the $\ell_{\max}$ value obtained as solution  of Eq. (\ref{criterio}) provides a quantitative criterium to know \textsl{a priori} whether
the power-law behavior in the autocorrelation function $C(\ell)$ of $\{z_i\}$ is observable or not. Note that a small $b_1$ value typically implies an also small $\ell_{\max}$, so that the
linear approximation becomes not significant for small $\ell$ values. In addition, the small $b_1$ value implies a poorly linear $C(C_G)$ function, so that in order to neglect higher order terms in the
expansion (\ref{linearC}), small $C_G(\ell)$ values are needed or, equivalently, large $\ell$ values. Therefore, when $b_1$ is small, the power-law behavior may be not observable since the $\ell$
values required can be larger than $\ell_{\max}$, where $C(\ell)$ is not significant.

\begin{figure}[h]
\includegraphics[width=8cm]{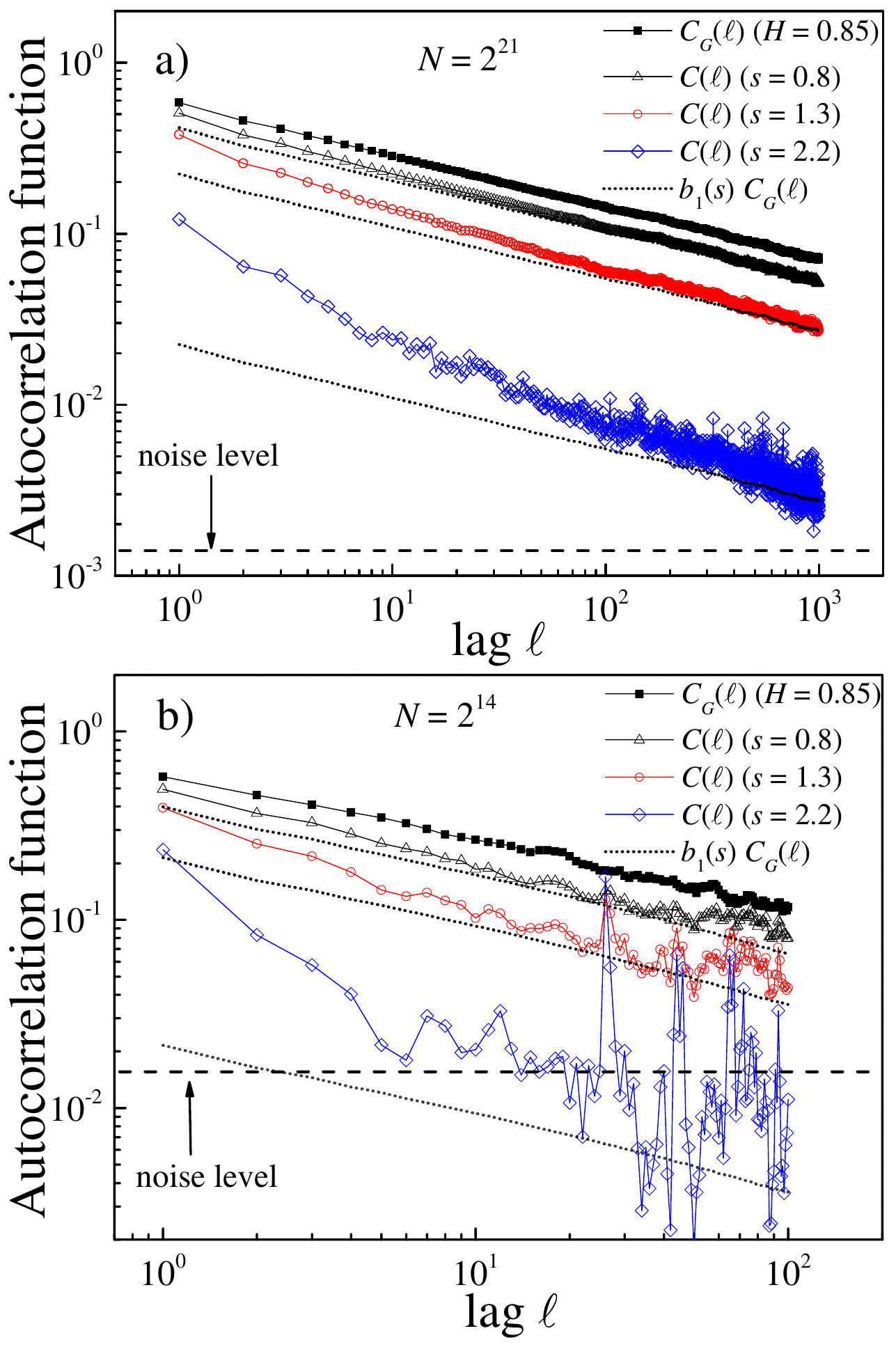}
\caption{(Color online) a) Autocorrelation function $C_G(\ell)$ for a Gaussian time series $\{z_{G,i}\}$ of length $N=2^{21}$ obtained using FFM with $H=0.85$. In the same panel, we show
the autocorrelation functions $C(\ell)$ of three lognormally distributed time series $\{z_i\}$ with different shape parameter $s$, which are obtained from $\{z_{G,i}\}$ using Eq. (\ref{transforma1}). The dotted lines correspond to the linear approximations $C(\ell)=b_1C_G(\ell)$, and the noise level is shown as an horizontal dashed line. b) The same as in panel a), but with $N=2^{14}$.} \label{auto-logn}
\end{figure}

The general rule is then that the power-law behavior of $C(\ell)$ is favoured to be observed when $\ell_{\max}$ is large enough, corresponding to have a $\{z_i\}$ time series with large size $N$
and/or with marginal distribution with large $b_1$ value. Obviously, a large $N$ value can compensate a small $b_1$ value and \textsl{viceversa}, but if both $N$ and $b_1$ are small then the power-law
behavior in $C(\ell)$ will not be observable. We illustrate these arguments in Figs. \ref{auto-logn}a) and b) where we consider times series with $N=2^{21}$ and $N=2^{14}$ respectively.
First, we show  the autocorrelation functions $C_G(\ell)$ of two Gaussian time series $\{z_{G,i}\}$ obtained via FFM with $H=0.85$. Then, using Eq. (\ref{transforma1}), each Gaussian time series is transformed into three lognormally distributed time series with different values of the shape parameter $s$ (0.8, 1.3 and 2.2), and the corresponding autocorrelation functions $C(\ell)$ are also shown  in Figs. \ref{auto-logn}a) and b).  We recall that the larger $s$, the smaller $b_1$ (see Fig. \ref{fig-logn}). In particular, the $b_1$ values are 0.714 for $s=0.8$, 0.382 for $s=1.3$, and $3.86\times 10^{-2}$ for $s=2.2$. Using these $b_1$ values, we also plot for each case the linear approximations (\ref{linearC}) in Figs. \ref{auto-logn}a) and \ref{auto-logn}b).

For the $N=2^{21}$ case (Fig. \ref{auto-logn}a)), we first observe that $C_G(\ell)$ behaves almost as a perfect power-law, $C_G(\ell)\propto \ell^{2H-2}=\ell^{-0.3}$,
in agreement with Eq. (\ref{cor-fgn}). For the lognormal time series, we find that the larger $b_1$, the smaller the $\ell$ value where the power-law behavior is reached, as expected. Indeed, for the case $s=0.8$, the corresponding lognormal time series exhibits power-law autocorrelation behavior practically in the whole $\ell$-range. For the intermediate $s=1.3$ value, the linear approximation requires a larger $\ell$ to be correct, and the power-law behavior of $C(\ell)$ happens at about $\ell \sim 100$. For the largest
$s=2.2$ value, $C(\ell)$ reaches the power-law behavior at larger $\ell$ values ($\ell\sim 1000$). In this $\ell$-range, $C(\ell)$ is noisier than in previous cases, since the values are close to the noise level, which in this case turns out to be $2/\sqrt{2^{21}}\simeq 1.4 \times 10^{-3}$, and which is also shown in Fig. \ref{auto-logn}a) as a horizontal dashed line. 

For the $N=2^{14}$ case (Fig. \ref{auto-logn}b)), the noise level (shown as a horizontal dashed line) is larger, around $2/\sqrt{2^{14}}\simeq 1.56 \times 10^{-2}$. As a consequence, although $C_G(\ell)$ exhibits the correct power-law behavior, $C_G(\ell) \sim \ell^{2H-2}=\ell^{-0.3}$, although a bit noisier than in Fig. \ref{auto-logn}a). For the lognormal times series, the observed behavior depends on the shape parameter $s$ value. For $s=0.8$, the corresponding $b_1$ value is large (0.72), and then the linear approximation is good enough to observe a power-law behavior of the corresponding $C(\ell)$.
Similarly, for the intermediate $s=1.3$ value, $b_1=0.382$ and the power-law behavior of $C(\ell)$ is also present for large $\ell$, but with higher noise around the linear approximation $b_1C_G(\ell)$. However, for $s=2.2$, the $b_1$ value is very small ($b_1=3.86\times 10^{-2}$) and then the linear approximation is never reached since before that happens, the $C(\ell)$ values are in the noise level range, and no power-law behavior is observed at all. In other words, in practice it is not possible to generate a power-law correlated lognormally distributed time series of length $N=2^{14}$ and shape parameter $s=2.2$. 

The behavior shown in Figs. \ref{auto-logn}a) and b) can be understood using the $\ell_{\max}$ solution of Eq. \ref{criterio}. Let us consider the worst case $s=2.2$ with $b_1=3.86\times 10^{-2}$. For the $N=2^{21}$ case,  we obtain $\ell_{\max}\simeq 11600$, large enough for $C(\ell)$ (diamonds in Fig. \ref{auto-logn}a)) to reach the linear approximation before entering into the noise level range. This is a case where the small $b_1$ value is compensated with a large series size $N$. However, for $N=2^{14}$ we  obtain $\ell_{\max}\simeq 3.64$, too small for $C(\ell)$  (diamonds int Fig. \ref{auto-logn}b)) to reach the validity region of the linear approximation before entering into the noise level range. 

Although we have used the lognormal distribution in the previous discussion, the conclusions are general: The controlled and observable power-law behavior of $C(\ell)$ is favoured for time series
$\{z_i\}$ following marginal distributions with large $b_1$ value, i.e, with very linear $C(C_G)$ functions. In addition, for a fixed distribution (fixed $b_1$), the larger the time series length $N$, the smaller the noise level, and the more likely to observe the power-law behavior of $C(\ell)$. Since the effect of the time series length $N$ is clear, we analyze the two properties of the marginal distribution of $\{z_i\}$ that, according to the results presented in Sec. \ref{distribuciones}, control the $b_1$ value: i) the tail behavior, and ii) the distribution symmetry. 

\begin{enumerate}
\item[(i)] Concerning the behavior of the tail of the distribution, we note that in general $b_1$ is large for bounded and for short, exponentially-bounded tail distributions. This is the case of the logistic ($b_1=0.99$), uniform ($b_1=0.95$), arcsine $(b_1=0.90)$, Laplace ($b_1=0.96$) and exponential ($b_1=0.81$) distributions. Note that all these $b_1$ values are larger than 0.72, which is the lognormal case shown in Fig. \ref{auto-logn} for $s=0.8$, and therefore the five corresponding autocorrelations functions will follow almost perfectly the linear approximation, and will behave
practically as perfect power-laws. But $b_1$ can also be large even for distributions with heavy tails, controlled by a shape parameter: the faster the decay of the heavy tail, the larger the corresponding $b_1$ value. This is the case of the symmetric Pareto (Fig. \ref{figparsim}b)), lognormal (Fig. \ref{fig-logn}b)), Weibull (Fig. \ref{fig-wei-par}c)) and Pareto (Fig. \ref{fig-wei-par}d))  distributions. Then, in general, we conclude that the faster the decay of the tail (even heavy) of the distribution of the $\{z_i\}$ time series, the larger the likelihood of observing a power-law behavior of $C(\ell)$, and viceversa.

\item[(ii)] Concerning the symmetry, symmetric distributions are in general better indicated to generate power-law correlated time series than non-symmetric distributions. The reason is that in the symmetric case, the expansion in Eq. ({\ref{serie2}}) only contains odd terms, as shown in (\ref{serie4}). Then, the first order approximation (\ref{linearC}) is more 
likely to be valid even for large $|C_G|$ values, or equivalently, for small $\ell$ values, than if the second order term is present, as it happens in non-symmetric distributions. We are aware that, since the behavior of the tail of the distribution also affects the $b_1$ value, one can have a non-symmetric short-tail distribution with a $b_1$ value larger than the one corresponding to a symmetric heavy-tail distribution. However, for symmetric and non-symmetric distributions with similar tail behavior, the $b_1$ value of the symmetric case is expected to be larger than for the non-symmetric one due to the absence of even terms in the expansion of the former. And indeed this is the case: for example, the non-symmetric exponential distribution and the symmetric Laplace distribution present identical exponential tail behavior, and the corresponding $b_1$ values are 0.81 and 0.96 respectively. As another example, for the symmetric Pareto and the Pareto distributions with the same value of shape parameter $\varepsilon$ controlling the power-law tail (see Tables I and II), the $b_1$ value for the symmetric case is always larger than for the non-symmetric one, as shown in Fig. \ref{fig-wei-par}d).   

\end{enumerate}

\subsection{Application II: modeling absolute returns in stock markets}

A well-known example of real-world time series  with autocorrelation function exhibiting power-law tails is the series of absolute returns of stock market prices \cite{podobnik}. 
Let us consider that $p(i)$ is the stock price at time $i$, where $i$ can be measured in minutes, hours, days, etc. The absolute return $r_i$ is defined as:
\begin{equation}
r_i\equiv \left| \log \left( \frac{p(i+1)}{p(i)} \right) \right|
\end{equation}

\begin{figure}[h]
\includegraphics[width=8cm]{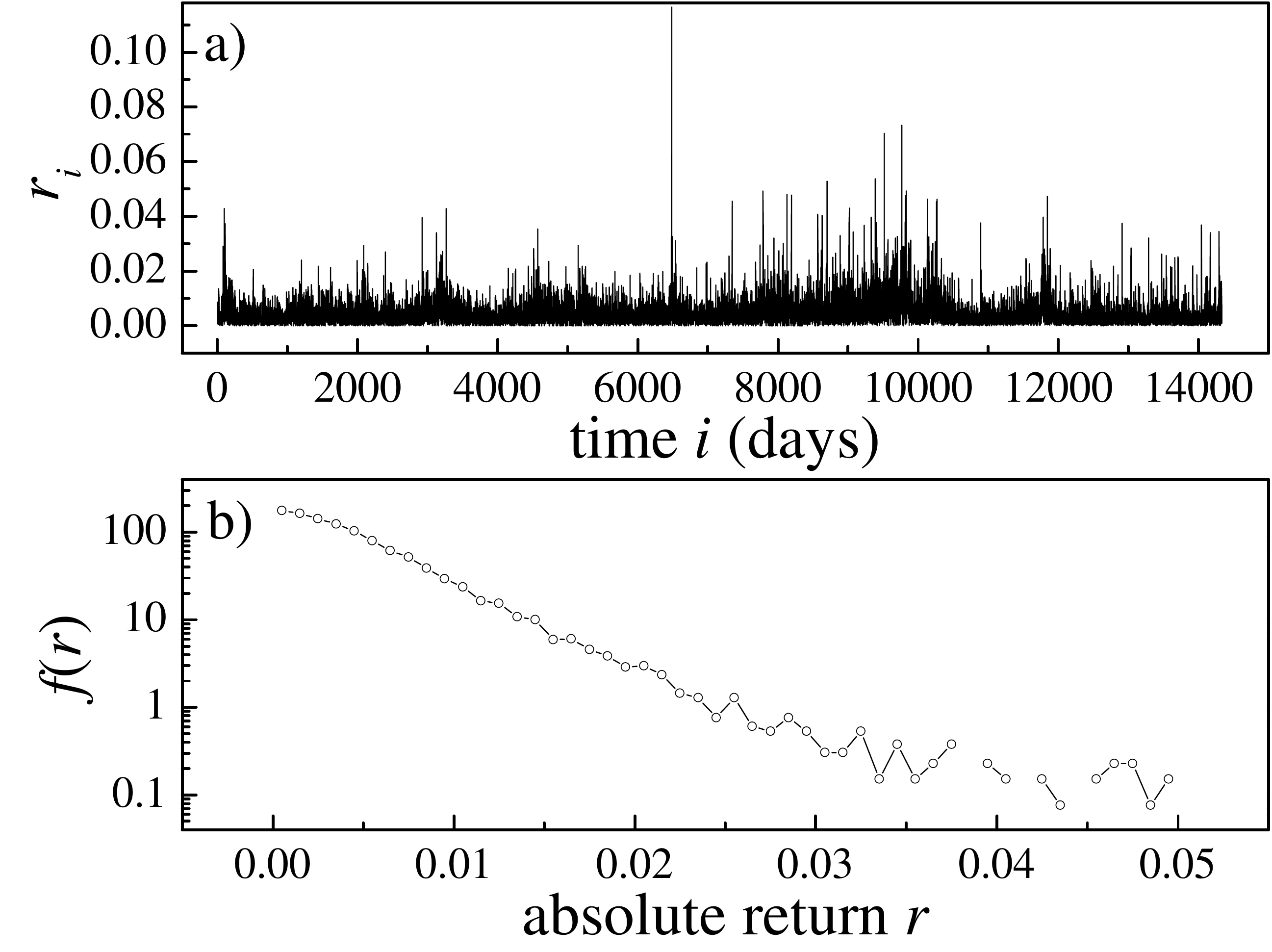}
\caption{a) Absolute returns time series $\{r_i\}$ of the IBM dayly stock price. Data cover since 1962. b) Probabilty density $f(r)$ of the time series shown in part a).} \label{IBM1}
\end{figure}

Typically, the time series $\{r_i\}$ present autocorrelation function with power-law tails, but the marginal distribution of $\{r_i\}$ is not Gaussian. As an example, we consider here
the daily absolute returns of IBM obtained from the NYSE, which are shown in Fig. \ref{IBM1}a) and the data cover the time range since 1962 with $N\simeq 14400$ data points. The marginal distribution of the data is not Gaussian,
as we show in Fig. \ref{IBM1}b) where we plot the probability density $f(r)$ obtained numerically. Since $f(r)$ is very linear using log-scale in the vertical axis, this indicates an almost exponential distribution, although with a heavier tail than exponential for large $r$ values.

\begin{figure}[h]
\includegraphics[width=8cm]{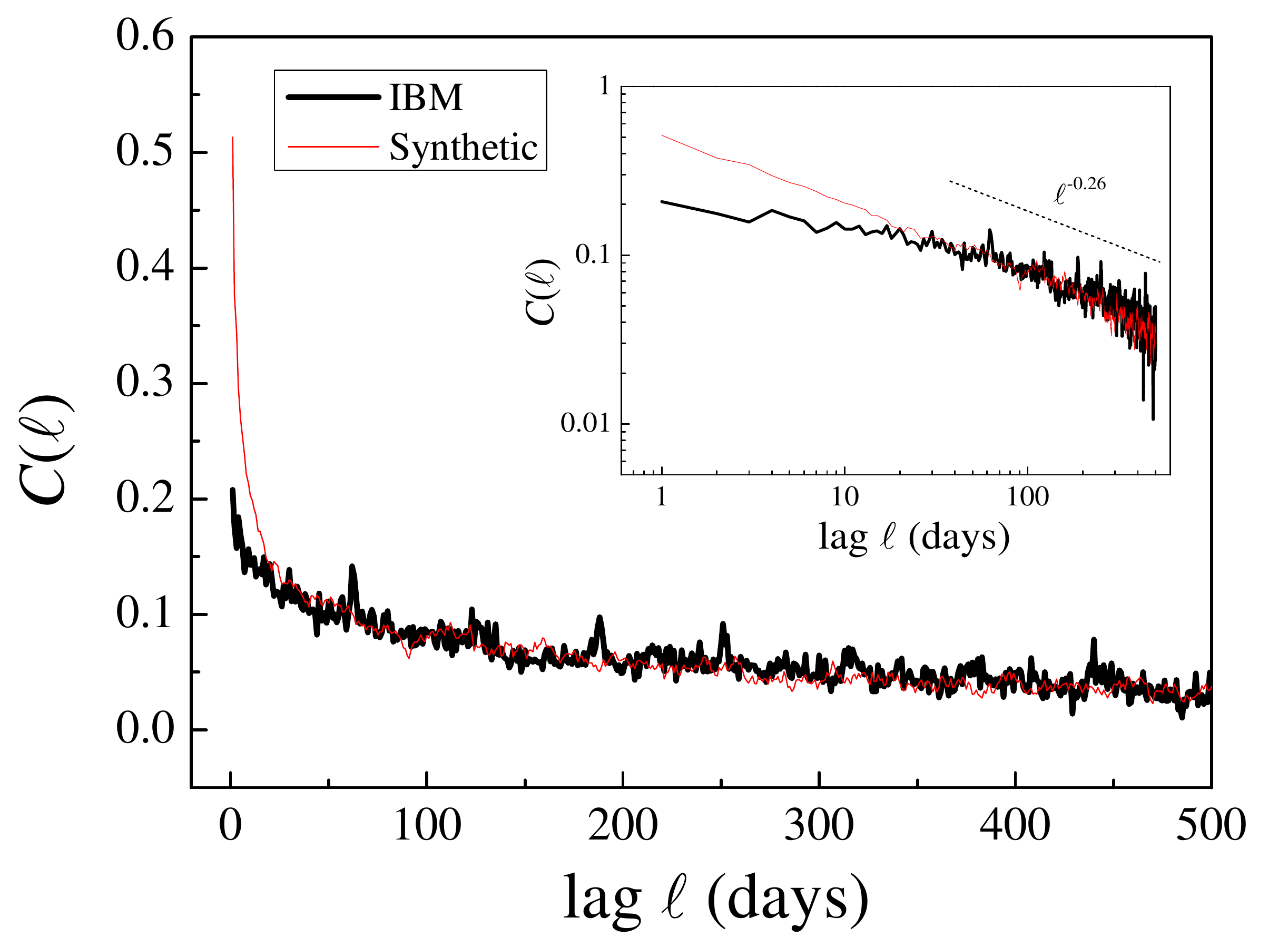}
\caption{Autocorrelation function of the IBM absolute returns shown in Fig. \ref{IBM1}a) (thick line). We also show in a thin line the autocorrelation function of a synthetic time series
obtained by generating a FFM Gaussian time series with $H=0.87$, which is then transformed using Eq. (\ref{transforma1}) to have the same marginal distribution as the IBM data. Inset: the two
autocorrelations shown in a log-log plot to better appreciate how both power-law tails match.} \label{IBM2}
\end{figure}

The autocorrelation function $C(\ell)$ of $\{r_i\}$ is shown in Fig. \ref{IBM2} (thick line). As shown in the inset, $C(\ell)$ presents a power-law tail of the form $C(\ell) \sim \ell^{-0.26}$.
Using the algorithm described above, we can generate a time series with the same power-law tail as the experimental data, and the same marginal distribution. To proceed, we first note that
according to Eq. (\ref{cor-fgn}), $2-2H=0.26$ so that  $H=0.87$. Then, we use FFM to generate a Gaussian ${\cal{N}}(0,1)$ time series $\{z_G(i)\}$ with $N=14400$ and $H=0.87$. Next, we obtain numerically 
the cumulative distribution $F$ of the experimental data and its inverse $F^{-1}$, and finally we construct a final time series $\{z_i\}$ using $z_i=F^{-1}[\Phi(z_{G,i})]$ with $i=1,2,\ldots 14400$.

By construction, $\{z_i\}$ presents the same marginal distribution of the experimental IBM absolute returns $\{r_i\}$. Using $F^{-1}$, we solve numerically the integral in Eq. (\ref{bn}) to obtain the $b_1$ value for the marginal distribution of $\{r_i\}$, and we get $b_1\simeq 0.643$, a quite large value, indicating that the linear approximation (\ref{linearC}) is good. In addition, by solving Eq. (\ref{criterio}) we obtain $\ell_{\max}\simeq 11000$. Both results imply that the power-law behavior of the autocorrelation function of $\{z_i\}$ with the correct exponent $H=0.87$ is reached for small $\ell$ values, and is significant practically in the whole range of $\ell$ (large $\ell_{\max}$). The autocorrelation function of $\{z_i\}$ is shown in Fig. \ref{IBM2} as a thin line. We note that  both autocorrelation functions present almost identical values in the whole range (up to 500 days) with discrepancies only for small lags. In the inset, we observe in a log-log plot how the autocorrelation function of the synthetic time series $\{z_i\}$ matches perfectly the power-law tail of the experimental $C(\ell)$.

\section{Conclusions}

Many real-world correlated time series are not Gaussian. However, very often the algorithms used to create surrogate time series produce correlated Gaussian time series with
prescribed autocorrelation function, which are then transformed to have the desired final marginal distribution. However, this last transformation always modify the Gaussian autocorrelation function.
In this work we have considered two stochastic Gaussian variables, $x_G$ and $y_G$, and we have transformed them respectively into two stochastic variables $x$ and $y$ following any arbitrary marginal
distribution. When the Gaussian variables are correlated with a given $C_G$ value, we have investigated how the correlation $C$ of the final variables $x$ and $y$ depends on $C_G$.
The function $C(C_G)$, which can be exactly determined by solving a 2D integral, turns out to depend on the properties of the destination distribution. We have obtained some general
properties of $C(C_G)$, such that $C(C_G)$ is an odd  function when the destination distribution is symmetric. In addition, we have obtained analytically a power expansion of $C(C_G)$,
which allows to weight the contribution of the different $C_G$ powers, and can be used to measure the linearity of $C(C_G)$ using the value of the first-order expansion coefficient $b_1$. We also have studied the specific behavior of $C(C_G)$ for several destination distributions with different properties concerning the support, the symmetry and the tail behavior. In general, destination distributions with bounded support present large $b_1$ values and therefore highly linear $C(C_G)$ functions. Also, the linearity of $C(C_G)$ is favoured for symmetric distributions, and for distributions with fast-decaying tails of exponential or faster than exponential type. $C(C_G)$ can behave also quite linearly even for heavy-tailed distributions of stretched-exponential or power-law type, but in general we observe that the longer the tail, the smaller the $b_1$  value and the linearity of $C(C_G)$. These results can be naturally extended to time series: 
when a Gaussian time series with autocorrelation $C_G(\ell)$ is transformed into another time series with arbitrary marginal distribution, the final series autocorrelation function $C(\ell)$ is determined by the $C(C_G)$ function of the destination distribution via $C(\ell)=C(C_G(\ell))$. In particular, for time series following marginal distributions with large $b_1$ values we have shown that $C(\ell) \simeq b_1 C_G(\ell)$. Using this property, we have extended the FFM algorithm, which produces Gaussian time series with a prescribed power-law autocorrelation function $C_G(\ell)$, to an algorithm
able to create time series with arbitrary marginal distribution and the same prescribed asymptotic power-law behavior as the Gaussian time series. We have used this algorithm to create a time series replicating both the marginal distribution and the autocorrelation power-law tail of a real-world time series: the absolute returns of a technological company.

\acknowledgements
We acknowledge financial support  by the Consejer\'{\i}a de Conocimiento, Investigaci\'on y Universidad, Junta de Andalucía  and European Regional Development Fund (ERDF), ref. SOMM17/6105/UGR and FQM-362.

\end{document}